\documentclass[12pt]{article}          
\usepackage{graphicx}

\usepackage{fullpage}
\usepackage{authblk}
\usepackage{txfonts}
\usepackage{fixmath}
\renewcommand{\baselinestretch}{1.66}\normalsize 

\renewenvironment{abstract}{\begin{quote} \bf}{\end{quote}}

\usepackage{setspace}
\usepackage{subfig}
\usepackage{booktabs}
\usepackage{multirow,rotating} 
\usepackage{color}
\usepackage{caption}

\def\spacecmd{} 
\def\spacecmdFig{}

\newcommand{\sciLet}[1]{\textbf{#1,}}
\newcommand{\SI}[0]{Supporting Information S1,}
\newcommand{\vrtLbl}[1]{\multirow{8}{1ex}{\begin{sideways}#1\end{sideways}}} 

\renewcommand{\thesubfigure}{\Alph{subfigure}} 

\newcommand{\event}[2][]{{#2}_\mathrm{event}{#1}}
\newcommand{\normal}[2][]{\left<{#2}_\mathrm{normal}{#1}\right>}
\newcommand{\avg}[1]{\left<{#1}\right>}
\def\s{\sigma}
\def\rc{r_\mathrm{c}}
\def\fmid{f_\mathrm{mid}}
\def\Pf{P}
\def\nrm{\mathrm{normal}}

\def\lastFigNum{5}
\date{March 1, 2011}

\begin{document}
	
\title{Collective response of human populations to large-scale emergencies}
\author[1,2,*]{James P.~Bagrow}
\author[1,2,*]{Dashun Wang}
\author[1,2,3]{Albert-L\'aszl\'o Barab\'asi}
\affil[1]{Center for Complex Network Research, Department of Physics, Northeastern University, Boston, MA 02115.}
\affil[2]{Center for Cancer Systems Biology, Dana-Farber Cancer Institute, Boston, MA 02115, USA.}
\affil[3]{Department of Medicine, Harvard Medical School, Boston, MA 02115, USA.}
\affil[*]{These authors contributed equally to this work.}
	
\maketitle 

\setstretch{2}

\begin{abstract}
    Despite recent advances in uncovering the quantitative features of stationary human activity
    patterns, many applications, from pandemic prediction to emergency response, require an
    understanding of how these patterns change when the population encounters unfamiliar conditions.
    To explore societal response to external perturbations we identified real-time changes in
    communication and mobility patterns in the vicinity of eight emergencies, such as bomb attacks
    and earthquakes, comparing these with eight non-emergencies, like concerts and sporting events.
    We find that communication spikes accompanying emergencies are both spatially and temporally
    localized, but information about emergencies spreads globally, resulting in communication
    avalanches that engage in a significant manner the social network of eyewitnesses. These results
    offer a quantitative view of behavioral changes in human activity under extreme conditions, with
    potential long-term impact on emergency detection and response. 
\end{abstract}


\section*{Introduction}
Current research on human dynamics is limited to data collected under normal and stationary
circumstances~\cite{vespScience2009}, capturing the regular daily activity of
individuals~\cite{Brockmann:nature2006,martaMobilityNature,ZtorNatureModelingDisease,LazerScience2009,chaomingLimitsScience,jpMobilePNAS,caldarelliInvasionJSTAT2009,DiegoRybski08042009,HerrmannAgentBasedPRE2009,brockmannEpidemicsPNAS2004,VittoriaColizza02142006,NoveltyAttentionHubermanPNAS,
    HumanDynamicsWebPRE,OnlinePopularityVespPRL}. Yet, there is exceptional need to understand how
people change their behavior when exposed to rapidly changing or unfamiliar
conditions~\cite{vespScience2009}, such as life-threatening epidemic
outbreaks~\cite{ZtorNatureModelingDisease,VittoriaColizza02142006}, emergencies and traffic
anomalies, as models based on stationary events are expected to break down under these
circumstances. Such rapid changes in conditions are often caused by natural, technological or
societal disasters, from hurricanes to violent conflicts~\cite{Bohorquez:2009p1440}. The
possibility to study such real time changes has emerged recently thanks to the widespread use of
mobile phones, which track both user
mobility~\cite{Brockmann:nature2006,martaMobilityNature,chaomingLimitsScience,DuyguBalcan12222009}
and real-time communications along the links of the underlying social
network~\cite{jpMobilePNAS,CaldarelliScaleFreeBook}. Here we take advantage of the fact that
mobile phones act as \emph{in situ} sensors at the site of an emergency, to study the real-time
behavioral patterns of the local population under external perturbations caused by emergencies.
Advances in this direction not only help redefine our understanding of information
propagation~\cite{DamonCentola09032010} and cooperative human actions under externally induced
perturbations, which is the main motivation of our work, but also offer a new perspective on
panic~\cite{Helbing:2000p1441,smallpoxModelKaplanPNAS2002,PhysRevLett.97.168001,buttsEmergentWTC2008}
and emergency protocols in a data-rich environment~\cite{eagleQuakePreprint}.

Our starting point is a country-wide mobile communications dataset, culled from the anonymized
billing records of approximately ten million mobile phone subscribers of a mobile company which
covers about one-fourth of subscribers in a country with close to full mobile penetration.  It
provides the time and duration of each mobile phone call~\cite{jpMobilePNAS}, together with
information on the tower that handled the call, thus capturing the real-time locations of the
users~\cite{martaMobilityNature,chaomingLimitsScience,Lambiotte20085317} (Methods, \SI{} Fig.~A).
To identify potential societal perturbations, we scanned media reports pertaining to the coverage
area between January 2007 and January 2009 and developed a corpus of times and locations for eight
societal, technological, and natural emergencies, ranging from bombings to a plane crash,
earthquakes, floods and storms (Table 1). Approximately 30\% of the events mentioned in the media
occurred in locations with sparse cellular coverage or during times when few users are active (like
very early in the morning). The remaining events do offer, however, a sufficiently diverse corpus to
explore the generic vs.~unique changes in the activity patterns in response to an emergency. Here we
discuss four events, chosen for their diversity: (1) a bombing, resulting in several injuries (no
fatalities); (2) a plane crash resulting in a significant number of fatalities; (3)  an earthquake
whose epicenter was outside our observation area but affected the observed population, causing mild
damage but no casualties; and (4) a power outage (blackout) affecting a major metropolitan area
(\SI{} Fig.~B).  To distinguish emergencies from other events that cause collective changes in human
activity, we also explored eight planned events, such as sports games and a popular local sports
race and several rock concerts. We discuss here in detail a cultural festival and a large pop music
concert as non-emergency references (Table 1, see also \SI{} Sec.~B). The characteristics of the
events not discussed here due to length limitations are provided in \SI{} Sec.~I for completeness
and comparison.

\section*{Results and Discussion}
\begin{figure*}\centering
	\subfloat{\label{fig:combinedTimeSeries:RawTimeSeries}} 
	\subfloat{\label{fig:combinedTimeSeries:normedTimes}}   
	\subfloat{\label{fig:combinedTimeSeries:changeNrho}}    
	\includegraphics[width=0.75\textwidth]{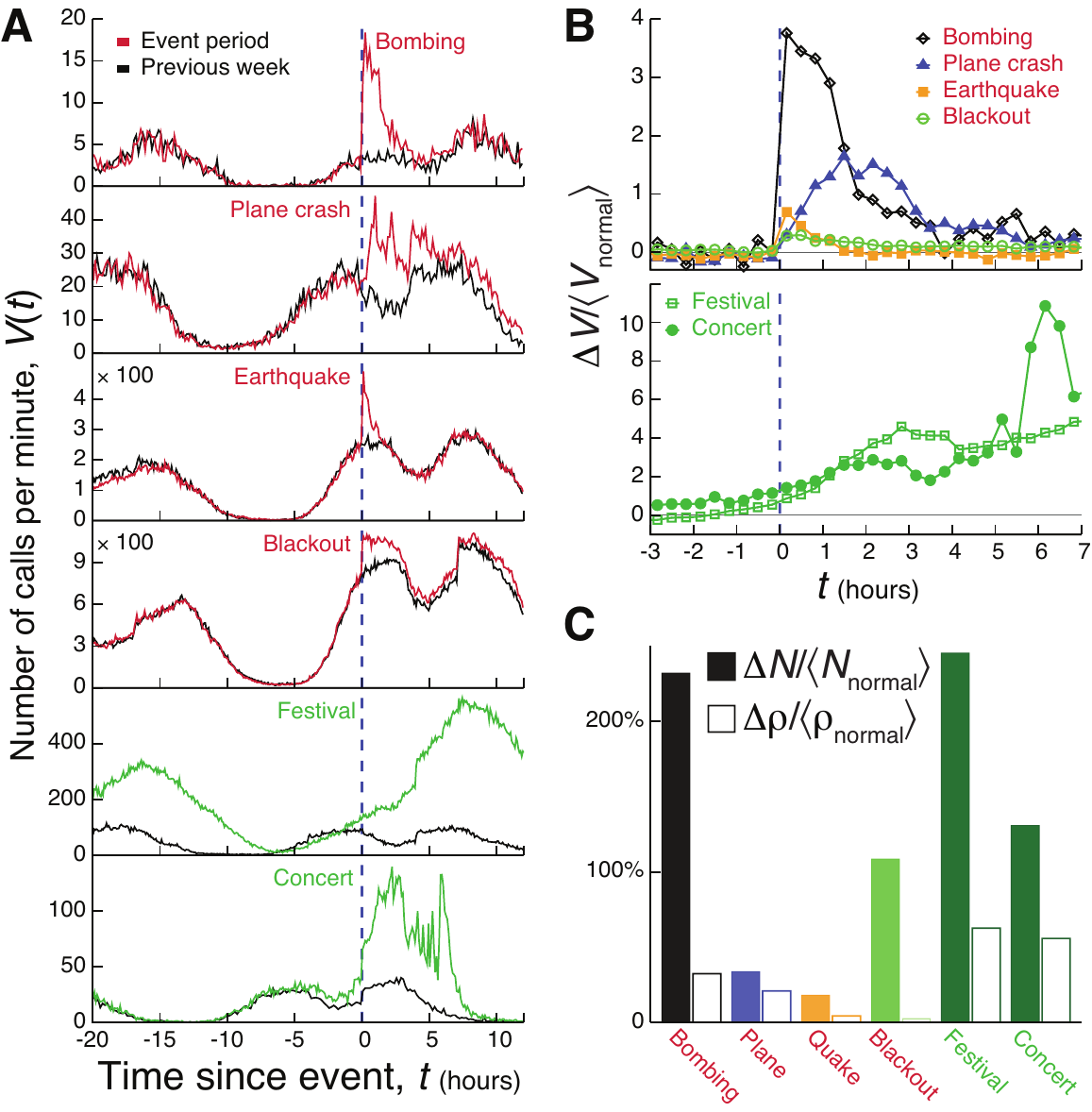}%
\caption*{\textbf{Call anomalies during emergencies.}
    \sciLet{\subref*{fig:combinedTimeSeries:RawTimeSeries}} The time dependence of call volume
    $V(t)$ in the vicinity of four emergencies and two non-emergencies (See Table 1).
    \sciLet{\subref*{fig:combinedTimeSeries:normedTimes}} The temporal behavior of the relative call
    volume $\Delta V/ \normal{V}$ of the events shown in \textbf{A}, where $\Delta V =
    \event{V}-\normal{V}$, $\event{V}$ is the call volume on the day of the event (shown in red in
    \textbf{A}), and $\normal{V}$ is the average call volume during the same period of the week (the
    call volume during the previous week is shown in black in \textbf{B}).
    \sciLet{\subref*{fig:combinedTimeSeries:changeNrho}} The relative change in the average number
    of calls placed per user ($\rho$) and the total number of users ($N$) making calls from the
    region indicates that the call anomaly is primarily due to a significant increase in the number
    of users that place calls during the events.
	\label{fig:combinedTimeSeries}%
}%
\end{figure*}

As shown in Fig.~\ref{fig:combinedTimeSeries:RawTimeSeries}, emergencies trigger a sharp spike in
call activity (number of outgoing calls and text messages) in the physical proximity of the event,
confirming that mobile phones act as sensitive local ``sociometers'' to external societal
perturbations. The call volume starts decaying immediately after the emergency, suggesting that the
urge to communicate is strongest right at the onset of the event. We see virtually no delay between
the onset of the event and the jump in call volume for events that were directly witnessed by the
local population, such as the bombing, the earthquake and the blackout.  Brief delay is observed
only for the plane crash, which took place in an unpopulated area and thus lacked eyewitnesses.  In
contrast, non-emergency events, like the festival and the concert in
Fig.~\ref{fig:combinedTimeSeries:RawTimeSeries}, display a gradual increase in call activity, a
noticeably different pattern from the ``jump-decay'' pattern observed for emergencies.  See also
\SI{} Figs.~I and J.

To compare the magnitude and duration of the observed call anomalies, in
Fig.~\ref{fig:combinedTimeSeries:normedTimes} we show the temporal evolution of the relative call
volume $\Delta V / \normal{V}$ as a function of time, where $\Delta V = \event{V} - \normal{V}$,
$\event{V}$ is the call activity during the event and $\normal{V}$ is the average call activity
during the same time period of the week. As Fig.~\ref{fig:combinedTimeSeries:normedTimes} indicates,
the magnitude of $\Delta V / \normal{V}$ correlates with our relative (and somewhat subjective)
sense of the event's potential severity and unexpectedness:  the bombing induces the largest change
in call activity, followed by the plane crash; whereas the collective reaction to the earthquake and
the blackout are somewhat weaker and comparable to each other.  While the relative change was also
significant for non-emergencies, the emergence of the call anomaly is rather gradual and spans seven
or more hours, in contrast with the jump-decay pattern lasting only three to five hours for
emergencies (Figs.~\ref{fig:combinedTimeSeries:normedTimes}, \SI{} Figs.~I and J).  As we show in
Fig.~\ref{fig:combinedTimeSeries:changeNrho} (see also \SI{} Sec.~C) the primary source of the
observed call anomaly is a sudden increase of calls by individuals who would normally not use their
phone during the emergency period, rather than increased call volume by those that are normally
active in the area.

\begin{figure*}\centering
	 \subfloat{\label{fig:spatialProps:bombMaps}}        
	 \subfloat{\label{fig:spatialProps:dVforSomeRs}}     
	 \subfloat{\label{fig:spatialProps:dVintegratedVsR}} 
	\includegraphics[width=\textwidth]{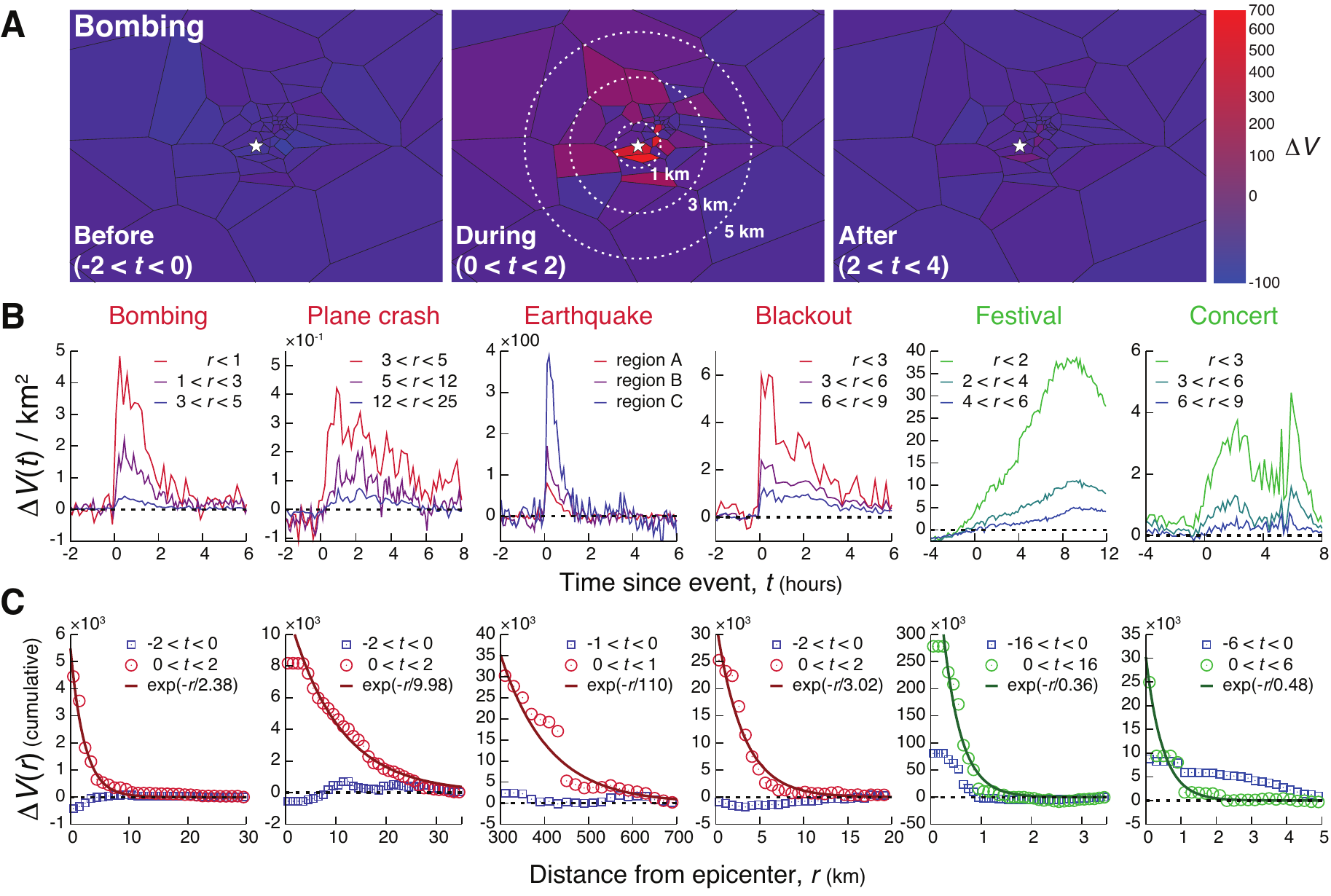}%
\caption*{\textbf{The spatial impact of an emergency.}
	\sciLet{\subref*{fig:spatialProps:bombMaps}} 
    Maps of total anomalous call activity (activity during the event minus expected normal activity)
    for two-hour periods before ($-2<t<0$), during ($0<t<2$), and after ($2<t<4$) the bombing. The
    color code corresponds to the total change $\sum_t \Delta V(t)$, where the sum runs over the
    particular time period.
	\sciLet{\subref*{fig:spatialProps:dVforSomeRs}} 
    Changes in call volume in regions at various distances $r$ from the event epicenter. Note that
    the peak of the call volume anomaly for the bombing within the observed $1<r<5$ km region is
    delayed by approximately 10 minutes compared to the $r<1$ km epicenter region. No call anomaly
    is observed for $r>10$ km.  The earthquake covers a large spatial range so we instead choose
    three event regions A-C, at distances of 310 km, 340 km, and 425 km from the seismic epicenter
    (which was outside the studied region).
	\sciLet{\subref*{fig:spatialProps:dVintegratedVsR}} 
    To measure the distance dependence of the anomaly, we computed the total anomalous call volume
    in \textbf{\subref*{fig:spatialProps:dVforSomeRs}} before ($\Delta t<t<0$) and after
    ($0<t<\Delta t$) each event as a function of the distance $r$, revealing approximately
    exponential decay, $\Delta V(r) \sim \exp\left(-r/\rc\right)$. Non-emergencies are spatially
    localized, with $\rc < 2$ km.
	\label{fig:spatialProps}%
}%
\end{figure*}

The temporally localized spike in call activity
(Fig.~\ref{fig:combinedTimeSeries:RawTimeSeries},\subref*{fig:combinedTimeSeries:normedTimes})
raises an important question: is information about the events limited to the immediate vicinity of
the emergency or do emergencies, often immediately covered by national media, lead to spatially
extended changes in call activity~\cite{buttsEmergentWTC2008}? We therefore inspected the change in
call activity in the vicinity of the epicenter, finding that for the bombing, for example, the
magnitude of the call anomaly is strongest near the event, and drops rapidly with the distance $r$
from the epicenter (Fig.~\ref{fig:spatialProps:bombMaps}).  To quantify this effect across all
emergencies, we integrated the call volume over time in concentric shells of radius $r$ centered on
the epicenter (Fig.~\ref{fig:spatialProps:dVforSomeRs}).  The decay is approximately exponential,
$\Delta V(r) \sim \exp\left(-r / \rc \right)$, allowing us to characterize the spatial extent of the
reaction with a decay rate $\rc$ (Fig.~\ref{fig:spatialProps:dVintegratedVsR}).  The observed decay
rates range from  $2$ km (bombing) to 10 km (plane crash), indicating that the anomalous call
activity is limited to the event's vicinity. An extended spatial range ($r_\mathrm{c}\approx 110$
km) is seen only for the earthquake, lacking a narrowly defined epicenter. Meanwhile, a
distinguishing pattern of non-emergencies is their highly localized nature: they are characterized
by a decay rate of less than $2$ km, implying that the call anomaly was narrowly confined to the
venue of the event.  This systematic split in $\rc$ between the spatially extended emergencies and
well-localized non-emergencies persists for all explored events (see Table 1, \SI{} Fig.~K).

\begin{figure*}\centering
	 \subfloat{\label{fig:incSocialDist:dVforSomeGi}}      
	 \subfloat{\label{fig:incSocialDist:dVintegratedVsGi}} 
	 \subfloat{\label{fig:incSocialDist:bombCascNet}}      
	 \subfloat{\label{fig:incSocialDist:bombPaths}}        
	\includegraphics[width=\textwidth]{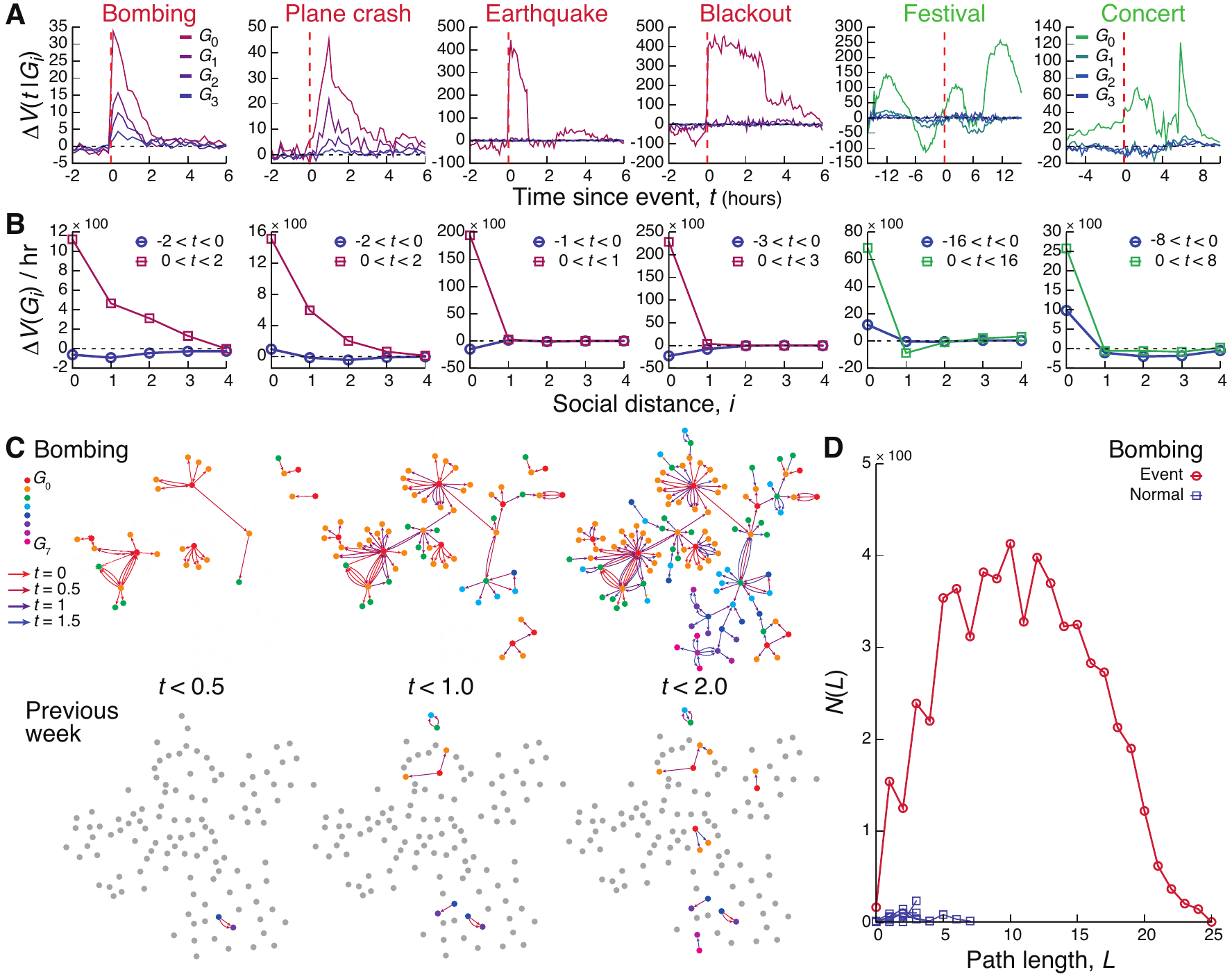}%
\caption*{\textbf{Social characteristics of information cascades.}
    \sciLet{\subref*{fig:incSocialDist:dVforSomeGi}} Changes in call volume for users directly
    affected by the event ($G_0$), users that receive calls from $G_0$ but are not near the event
    ($G_1$), users contacted by $G_1$ but not in $G_1$ or $G_0$ ($G_2$), etc.  For the bombing and
    plane crash, populations respond very rapidly, within minutes.
    \sciLet{\subref*{fig:incSocialDist:dVintegratedVsGi}} The total amount of anomalous call
    activity in \textbf{\subref*{fig:incSocialDist:dVforSomeGi}} before (during $-\Delta t < t < 0$)
    and after (during $0 < t < \Delta t$) the event for each user group $G_i$ quantifies the impact
    on the social network.  We see that information propagates deeply into the social network for
    the bombing and plane crash.
    \sciLet{\subref*{fig:incSocialDist:bombCascNet}} Top panel: the contact network formed between
    affected users during the bombing.  Bottom panel: the call pattern between users that are active
    during the emergency during the previous week, indicating that the information cascade observed
    during the bombing is out of the ordinary.
    \sciLet{\subref*{fig:incSocialDist:bombPaths}}  The distribution of shortest paths within the
    contact network quantifies the anomalous information cascade induced by the bombing.
	\label{fig:incSocialDist}%
}%
\end{figure*}

Despite the clear temporal and spatial localization of anomalous call activity during emergencies,
one expects some degree of information propagation beyond the eyewitness
population~\cite{rileyCraneYoutubePNAS2008}. We therefore identified the individuals located within
the event region $\left(G_0\right)$, as well as a $G_1$ group consisting of individuals outside the
event region but who receive calls from the $G_0$ group during the event, a $G_2$ group that receive
calls from $G_1$, and so on. We see that the $G_0$ individuals engage their social network within
minutes, and that the $G_1$, $G_2$, and occasionally even the $G_3$ group show an anomalous call
pattern immediately after the anomaly (Fig.~\ref{fig:incSocialDist:dVforSomeGi}). This effect is
quantified in Fig.~\ref{fig:incSocialDist:dVintegratedVsGi}, where we show the increase in call
volume for each group as a function of their social network based distance from the epicenter (for
example, the social distance of the $G_2$ group is 2, being two links away from the $G_0$ group),
indicating that the bombing and plane crash show strong, immediate social propagation up to the
third and second neighbors of the eyewitness $G_0$ population, respectively.  The earthquake and
blackout, less threatening emergencies, show little propagation beyond the immediate social links of
$G_0$ and social propagation is virtually absent in non-emergencies.

The nature of the information cascade behind the results shown in
Fig.~\ref{fig:incSocialDist:dVforSomeGi},\subref*{fig:incSocialDist:dVintegratedVsGi} is illustrated
in Fig.~\ref{fig:incSocialDist:bombCascNet}, where we show the individual calls between users active
during the bombing. In contrast with the information cascade triggered by the emergencies witnessed
by the $G_0$ users, there are practically no calls between the same individuals during the previous
week. To quantify the magnitude of the information cascade we measured the length of the paths
emanating from the $G_0$ users, finding them to be considerably longer during the emergency
(Fig.~\ref{fig:incSocialDist:bombPaths}), compared to five non-emergency periods, demonstrating that
the information cascade penetrates deep into the social network, a pattern that is absent during
normal activity~\cite{JPemergence}.  
See also \SI{} Figs.~E, F, G, H, L, M, N, and O, and Table A.

The existence of such prominent information cascades raises tantalizing questions about who
contributes to information propagation about the emergency.  Using self-reported gender information
available for most users (see Supporting Information S1), we find that during emergencies female
users are more likely to make a call than expected based on their normal call patterns. This gender
discrepancy holds for the $G_0$ (eyewitness) and $G_1$ groups, but is absent for non-emergency
events (see \SI{} Sec.~E, Fig.~C).  We also separated the total call activity of $G_0$ and $G_1$
individuals into voice and text messages (including SMS and MMS). For most events (the earthquake
and blackout being the only exceptions), the voice/text ratios follow the normal patterns (\SI{}
Fig.~D), indicating that users continue to rely on their preferred means of communication during an
emergency.

\begin{figure*}\centering
	\subfloat{\label{fig:bomb2:temp}} 
	\subfloat{\label{fig:bomb2:sptl}} 
	\subfloat{\label{fig:bomb2:socT}} 
	\subfloat{\label{fig:bomb2:SocI}} 
	\includegraphics[width=\textwidth]{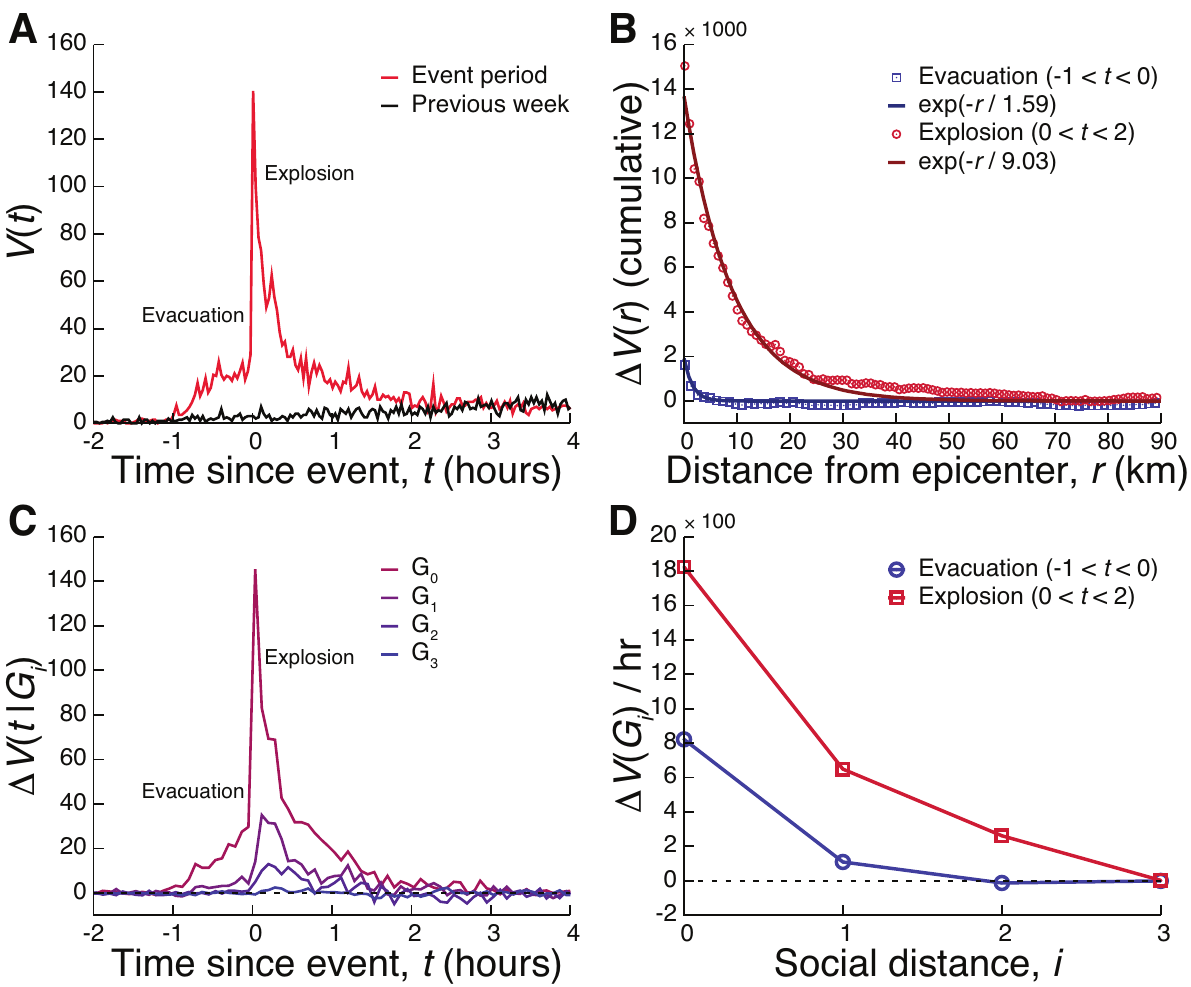}%
\caption*{\textbf{Analyzing a composite event (evacuation preceding an explosion).}
	\sciLet{\subref*{fig:bomb2:temp}}
    Call activity increases during the evacuation ($-1 < t < 0$) but levels off after the initial
    warning, until the explosion at $t=0$ causes a much larger increase in call activity.
	\sciLet{\subref*{fig:bomb2:sptl}}
    Spatially, the evacuation causes a sharply localized activity spike ($\rc = 1.6$ km), but the
    explosion increases the spatial extent dramatically ($\rc = 9.0$ km).
	\sciLet{\subref*{fig:bomb2:socT}--\subref*{fig:bomb2:SocI}}
    The evacuation only activates the $G_0$ (eyewitness) and $G_1$ groups, meaning that information
    fails to propagate significantly beyond the initial group and their immediate ties.  However,
    the blast not only leads to a further increase in call activity in the $G_0$ and $G_1$ groups,
    but also triggers the second neighbors $G_2$.
	\label{fig:bomb2}%
}
\end{figure*}

The patterns identified discussed above allow us to dissect complex events, such as an explosion in
an urban area preceded by an evacuation starting approximately one hour before the blast.  While a
call volume anomaly emerges right at the start of the evacuation, it levels off and the jump-decay
pattern characteristic of an emergency does not appear until the real explosion
(Fig.~\ref{fig:bomb2:temp}). The spatial extent of the evacuation response is significantly smaller
than the one observed during the event ($\rc=1.6$ for the evacuation compared with $\rc = 9.0$ for
the explosion, see Fig.~\ref{fig:bomb2:sptl}). During the evacuation, social propagation is limited
to the $G_0$ and $G_1$ groups only (Fig.~\ref{fig:bomb2:socT},\subref*{fig:bomb2:SocI}) while after
the explosion we observe a communication cascade that activates the $G_2$ users as well.  The lack
of strong propagation during evacuation indicates that individuals tend to be reactive rather than
proactive and that a real emergency is necessary to initiate a communication cascade that
effectively spreads emergency information.

The results of Figs.~\ref{fig:combinedTimeSeries}-\ref{fig:bomb2} not only
indicate that the collective response of the population to an emergency follows
reproducible patterns common across diverse events, but they also document
subtle differences between emergencies and non-emergencies.  We therefore
identified four variables that take different characteristic values for
emergencies and non-emergencies: (i) the midpoint fraction
$\fmid=\left(t_\mathrm{mid}-t_\mathrm{start}\right)/\left(t_\mathrm{stop}-t_\mathrm{start}\right)$,
where $t_\mathrm{start}$ and $t_\mathrm{stop}$ are the times when the anomalous
activity begins and ends, respectively, and $t_\mathrm{mid}$ is the time when
half of the total anomalous call volume has occurred; 
(ii) the spatial decay rate $\rc$ capturing the extent of the event;
(iii) the relative size $R$ of each information cascade, representing the ratio
between the number of users in the event cascade and the cascade tracked during
normal periods;
(iv) the probability for users to contact existing friends (instead of placing
calls to strangers).

\begin{figure*}\centering
	\subfloat{\label{fig:summaryProps:temprl_fmid}} 
	\subfloat{\label{fig:summaryProps:sptial_rc}}   
	\subfloat{\label{fig:summaryProps:social_R}}    
	\subfloat{\label{fig:summaryProps:social_zF}}   
	\includegraphics[width=0.75\textwidth]{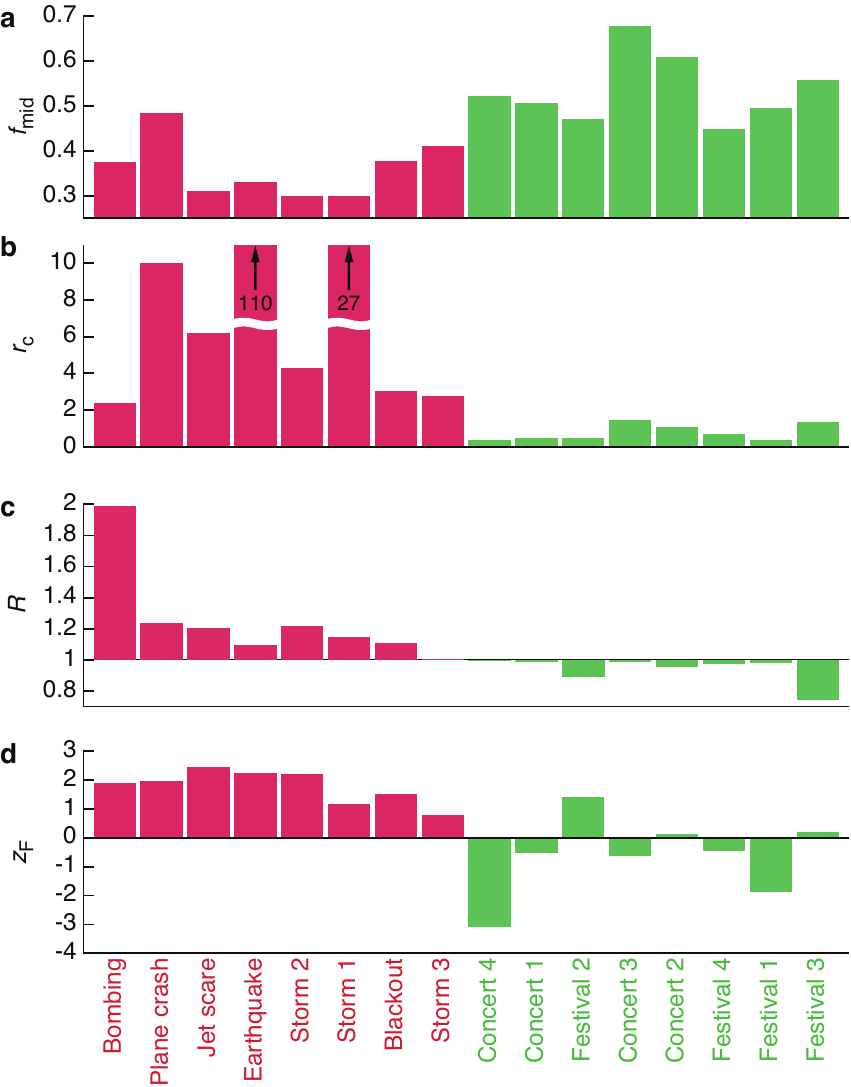}%
\caption*[Systematic response mechanisms during emergencies]{\textbf{Systematic response mechanisms during emergencies.}
	\sciLet{\subref*{fig:summaryProps:temprl_fmid}}
    The midpoint fraction $\fmid$ quantifying the onset speed of anomalous call activity (a lower
    $\fmid$ indicates a faster onset).  Emergencies display a more abrupt call anomaly than
    non-emergencies, which feature gradual buildups of anomalous call activity.
	\sciLet{\subref*{fig:summaryProps:sptial_rc}}
    The spatial extent of the events, quantified by $\rc$, indicates that non-emergency events are
    far more centrally localized than unexpected emergencies. 
	\sciLet{\subref*{fig:summaryProps:social_R}}
    The relative cascade size $R=\event{N}/\normal{N}$, where $N=\sum_i|G_i|$ is the number of users
    in the social cascade. 
	\sciLet{\subref*{fig:summaryProps:social_zF}}
    $z_\mathrm{F}=\left(\event{\Pf}-\normal{\Pf}\right)/\s\left(\Pf_\nrm\right)$, where $\Pf$ is the
    probability of calling an acquaintance and $\s(\Pf)$ is the standard deviation of $\Pf$.
	\label{fig:summaryProps}%
}
\end{figure*}

In Fig.~\ref{fig:summaryProps} we show these variables for all 16 events,
finding systematic differences between emergencies and non-emergencies.  As the
figure indicates, a multidimensional variable, relying on the documented changes
in human activity, can be used to automatically distinguish emergency situations
from non-emergency induced anomalies. Such a variable could also help real-time
monitoring of emergencies~\cite{eagleQuakePreprint}, from information about the
size of the affected population, to the timeline of the events, and could help
identify mobile phone users capable of offering immediate, actionable
information, potentially aiding search and rescue.

Rapidly-evolving events such as those studied throughout this work require dynamical data with
ultra-high temporal and spatial resolution and high coverage.  Although the populations affected by
emergencies are quite large, occasionally reaching thousands of users, due to the demonstrated
localized nature of the anomaly, this size is still small in comparison to other proxy studies of
human dynamics, which can exploit the activity patterns of millions of internet users or
webpages~\cite{NoveltyAttentionHubermanPNAS,
    HumanDynamicsWebPRE,OnlinePopularityVespPRL,JPemergence}. Meanwhile, emergencies occur over very
short timespans, a few hours at most, whereas much current work on human dynamics relies on
longitudinal datasets covering months or even years of activity for the same users
(e.g.~\cite{martaMobilityNature,chaomingLimitsScience,DiegoRybski08042009}), integrating out
transient events and noise.  But in the case of emergencies, such transient events are precisely
what we wish to quantify.  Given the short duration and spatially localized nature of these events,
it is vital to have extremely high coverage of the entire system, to maximize the availability of
critical information during an event. To push human dynamics research into such fast-moving events
requires new tools and datasets capable of extracting signals from limited data.  We believe that
our research offers a first step in this direction.

In summary, similar to how biologists use drugs to perturb the state of a cell to better understand
the collective behavior of living systems, we used emergencies as external societal perturbations,
helping us uncover generic changes in the spatial, temporal and social activity patterns of the
human population.
Starting from a large-scale, country-wide mobile phone dataset, we used news reports to gather a
corpus of sixteen major events, eight unplanned emergencies and eight scheduled activities. Studying
the call activity patterns of users in the vicinity of these events, we found that unusual activity
rapidly spikes for emergencies in contrast with non-emergencies induced anomalies that build up
gradually before the event; that the call patterns during emergencies are exponentially localized
regardless of event details; and that affected users will only invoke the social network to
propagate information under the most extreme circumstances. When this social propagation does occur,
however, it takes place in a very rapid and efficient manner, so that users three or even four
degrees from eyewitnesses can learn of the emergency within minutes.

These results not only deepen our fundamental understanding of human dynamics, but could also
improve emergency response. Indeed, while aid organizations increasingly use the distributed,
real-time communication tools of the 21st century, much disaster research continues to rely on
low-throughput, post-event data, such as questionnaires, eyewitness
reports~\cite{virginaTechShootingCellSurveys,QuarantelliHandbook2007}, and communication records
between first responders or relief organizations~\cite{buttsBrokerageKatrina2008}. The emergency
situations explored here indicate that, thanks to the pervasive use of mobile phones, collective
changes in human activity patterns can be captured in an objective manner, even at surprisingly
short time-scales, opening a new window on this neglected chapter of human dynamics.

\section*{Materials and Methods}
\subsection*{Dataset}
We use a set of anonymized billing records from a western european mobile phone service
provider~\cite{jpMobilePNAS,martaMobilityNature,chaomingLimitsScience}.  The records cover
approximately 10M subscribers within a single country over 3 years of activity.  Each billing
record, for voice and text services, contains the unique identifiers of the caller placing the call
and the callee receiving the call; an identifier for the cellular antenna (tower) that handled the
call; and the date and time when the call was placed.  Coupled with a dataset describing the
locations (latitude and longitude) of cellular towers, we have the approximate location of the
caller when placing the call. 
For full details, see \SI{} Sec.~A.  

\subsection*{Identifying events}
To find an event in the mobile phone data, we need to determine its time and location.  We have used
online news aggregators, particularly the local \texttt{news.google.com} service to search for news
stories covering the country and time frame of the dataset.  Keywords such as `storm', `emergency',
`concert', etc.~were used to find potential news stories.  Important events such as bombings and
earthquakes are prominently covered in the media and are easy to find.  Study of these reports,
which often included photographs of the affected area, typically yields precise times and locations
for the events. Reports would occasionally conflict about specific details, but this was rare.  We
take the \emph{reported} start time of the event as $t=0$.  

To identify the beginning and ending of an event, $t_\mathrm{start}$ and $t_\mathrm{stop}$, we
adopt the following procedure.  First, identify the event region (a rough estimate is sufficient)
and scan all its calls during a large time period covering the event (e.g., a full day), giving
$\event[(t)]{V}$.  Then, scan calls for a number of ``normal'' periods, those modulo one week from
the event period, exploiting the weekly periodicity of $V(t)$.  These normal periods' time series
are averaged to give $\normal{V}$.  (To smooth  time series, we typically bin them into 5--10 minute
intervals.)  The standard deviation $\s\left(V_\mathrm{normal}\right)$ as a function of time is then
used to compute $z(t) = \Delta V(t) / \s\left(V_\mathrm{normal}\right)$.  Finally, we define the
interval $\left(t_\mathrm{start}, t_\mathrm{stop}\right)$ as the longest contiguous run of time
intervals where $z(t)>z_\mathrm{thr}$, for some fixed cutoff $z_\mathrm{thr}$. We chose
$z_\mathrm{thr} = 1.5$ for all events. 

For full details, see \SI{} Sec.~B.

\section*{Acknowledgments}
The authors thank A.~Pawling, F.~Simini, M.~C.~Gonz\'alez, S.~Lehmann, R.~Menezes, N.~Blumm,
C.~Song, J.~P.~Huang, Y.-Y. Ahn, P.~Wang, R.~Crane, D.~Sornette, and D.~Lazer for many useful
discussions. 

\singlespacing

\renewcommand{\baselinestretch}{1.66}\normalsize

\clearpage

\begin{table}\centering
	\caption{\textbf{Summary of the studied emergencies and non-emergencies.}
    The columns provide the duration of the anomalous call activity (Fig.~1), the spatial decay rate
    $\rc$ (Fig.~2), the number of  users in the event population $|G_0|$, and the total size of the
    information cascade $\sum_i |G_i|$ (Fig.~3).  Events discussed in the main text are italicized,
    the rest are discussed in the supplementary material.  `Jet scare' refers to a sonic boom
    interpreted by the local population and initial media reports as an explosion.}
\renewcommand{\baselinestretch}{1}\normalsize
\begin{tabular}{llllllll}
\toprule 
                                      &    & Event 				 & duration (hours) & $\rc$ (km)	& $|G_0|$	& $\sum_i |G_i|$\\ %
\midrule
\vrtLbl{\bf \color{red}Emergencies}   & 1  & \textit{Bombing}	 & 1.92				& 2.38	   		& 750     	& 5,099  		\\ 
                                      & 2  & \textit{Plane crash}& 2.17				& 9.98			& 2,104    	& 7,325  		\\ 
                                      & 3  & \textit{Earthquake} & 1.42				& 110			& 32,403   	& 83,280 		\\ 
                                      & 4  & \textit{Blackout}	 & 3.0			 	& 3.02			& 84,751   	& 288,332		\\ 
                                      & 5  & Jet scare  		 & 1.67 			& 6.18			& 3,556 	& 11,575 		\\ 
                                      & 6  & Storm 1			 & 2.33				& 27.0			& 7,350  	& 18,124 		\\ 
                                      & 7  & Storm 2			 & 2.0				& 4.29			& 14,634 	& 33,963 		\\ 
                                      & 8  & Storm 3			 & 1.75				& 2.79			& 19,239	& 48,626 		\\ 
\midrule
\vrtLbl{\bf \color{green}Non-emergencies} & 9  & \textit{Concert 1}	 & 13.25			& 0.48			& 11,376 	& 91,889 		\\ 
                                      & 10 & Concert 2			 & 6.67 			& 1.06			& 3,939  	& 29,837 		\\ 
                                      & 11 & Concert 3			 & 9.08				& 1.48			& 5,134  	& 81,125 		\\ 
                                      & 12 & Concert 4			 & 12.08			& 0.35			& 2,630  	& 17,998 		\\ 
                                      & 13 & \textit{Festival 1} & 19.92			& 0.36			& 66,869 	& 454,687		\\ 
                                      & 14 & Festival 2			 & 2.17				& 0.50			& 1,453  	& 7,963  		\\ 
                                      & 15 & Festival 3			 & 20.92			& 1.33			& 10,854 	& 427,839		\\ 
                                      & 16 & Festival 4			 & 11.25			& 0.72			& 3,117  	& 16,822 		\\
\bottomrule
\end{tabular}
\end{table}

\clearpage

\renewcommand{\baselinestretch}{1.0}\normalsize

\renewcommand{\thefigure}{\Alph{figure}}
\renewcommand{\thesection}{\Alph{section}}
\renewcommand{\thetable}{\Alph{table}}
\renewcommand{\theequation}{S\arabic{equation}}
\renewcommand{\thepage}{S\arabic{page}}
\renewcommand{\thesubfigure}{\Alph{subfigure}} 

\setcounter{page}{1}

\spacecmd
\noindent{\huge Supporting Information}
\baselineskip24pt
\singlespacing
\noindent\textit{Collective response of human populations to large-scale emergencies} \\
\noindent by James P.~Bagrow, Dashun Wang, and  Albert-L\'aszl\'o Barab\'asi

\renewcommand*\contentsname{Table of Contents}
\tableofcontents
\listoffigures
\listoftables

\section{Dataset}\label{sec:dataset}
We use a set of anonymized billing records from a western european mobile phone service
provider~\cite{jpMobilePNAS,martaMobilityNature,chaomingLimitsScience}.  The records cover
approximately 10M subscribers within a single country over 3 years of activity.  Each billing
record, for voice and text services, contains the unique identifiers of the caller placing the call
and the callee receiving the call; an identifier for the cellular antenna (tower) that handled the
call; and the date and time when the call was placed.  Coupled with a dataset describing the
locations (latitude and longitude) of cellular towers, we have the approximate location of the
caller when placing the call. Unless otherwise noted, a ``call'' can be either voice or text (SMS,
MMS, etc.), and ``call volume'' or ``call activity'' is both voice calls and text messages. 

After identifying the start time and location of an event, we then scan these billing records to
determine the activity of nearby users.  The mobile phone activity patterns of the affected users
can then be followed in the weeks preceding or following the event, to provide control or baseline
behavior.

Self-reported gender information is available for approximately 90\% of subscribers.

\subsection{Market share}
These records cover approximately 20\% of the country's mobile phone market.  However, we also
possess identification numbers for phones that are outside the provider but that make or receive
calls to users within the company.  While we do not possess any other information about these lines,
nor anything about their users or calls that are made to other numbers outside the service provider,
we do have records pertaining to all calls placed to or from these ID numbers involving subscribers
covered by our dataset.  This information was used to study social propagation (see
Sec.~\ref{sec:calcSocProp:contactNet}).

\section{Identifying events}\label{sec:idevents}

To find an event in the mobile phone data, we need to determine its time and location.  We have used
online news aggregators, particularly the local \texttt{news.google.com} service to search for news
stories covering the country and time frame of the dataset.  Keywords such as `storm', `emergency',
`concert', etc.~were used to find potential news stories.  Important events such as bombings and
earthquakes are prominently covered in the media and are easy to find.  Study of these reports,
which often included photographs of the affected area, typically yields precise times and locations
for the events. Reports would occasionally conflict about specific details, but this was rare.  We
take the \emph{reported} start time of the event as $t=0$.

\begin{figure}[t]\centering
	\includegraphics[]{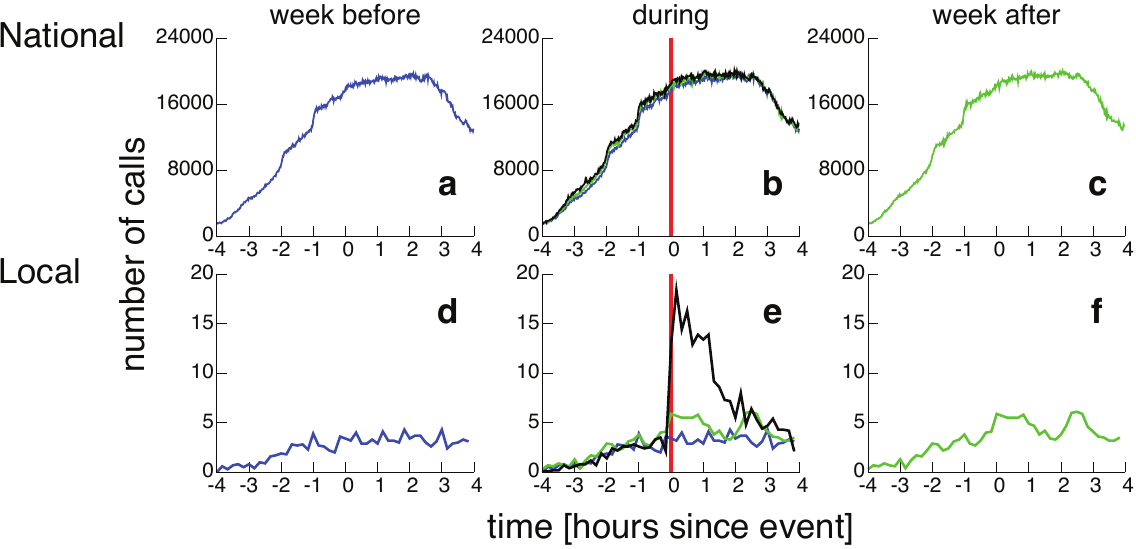}
	\caption[Regional and national visibility of events]{\spacecmdFig{} \small 
    \textbf{Regional and national visibility of events.} On a national level
    (\textbf{a}--\textbf{c}), the spike in call activity due to the bombing is lost, but it clearly
    emerges when we focus only on the immediately local vicinity of the event
    (\textbf{d}--\textbf{f}). The strong weekly periodicity in $V(t)$ is also visible.
	\label{fig:event_local_national}}
\end{figure}

Most events are spatially localized, so it is important to consider only the immediate event region.
Otherwise, the event signal is masked by normal activity (Fig.~\ref{fig:event_local_national}).  The
local event region can be estimated as a circle centered on the identified epicenter with a radius
chosen based on $\rc$.  For the blackout, however, we chose all towers within the affected city
(contained within the city's postal codes).

To identify the beginning and the end of an event, $t_\mathrm{start}$ and $t_\mathrm{stop}$, we
adopt the following procedure.  First, identify the event region (a rough estimate is sufficient)
and scan all its calls during a large time period covering the event (e.g., a full day), giving
$\event[(t)]{V}$.  Then, scan calls for a number of ``normal'' periods, those modulo one week from
the event period, exploiting the weekly periodicity of $V(t)$.  These normal periods' time series
are averaged to give $\normal{V}$.  (To smooth  time series, we typically bin them into 5--10 minute
intervals.)  The standard deviation $\s\left(V_\mathrm{normal}\right)$ as a function of time is then
used to compute $z(t) = \Delta V(t) / \s\left(V_\mathrm{normal}\right)$.  Finally, we define the
interval $\left(t_\mathrm{start}, t_\mathrm{stop}\right)$ as the longest contiguous run of time
intervals where $z(t)>z_\mathrm{thr}$, for some fixed cutoff $z_\mathrm{thr}$. We chose
$z_\mathrm{thr} = 1.5$ for all events.

\begin{figure}[t]\centering
	\includegraphics[]{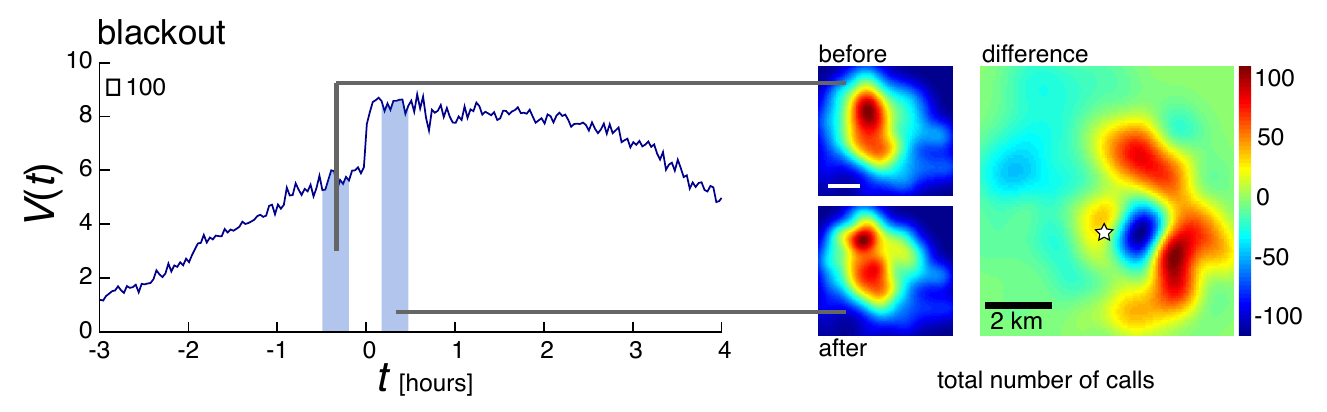}
	\caption[Spatial changes in call activity due to the blackout]{\spacecmdFig{} \small 
	\textbf{Spatial changes in call activity due to the blackout.} 
    We integrate call activity $V(t)$ over two time windows before and after the blackout occurs
    (\textbf{left}, shaded).  Studying total call load spatially we see a nonlinear response, with a
    region near the city center (star) suffering a drop in calls due to the blackout
    (\textbf{right}).  This region is surrounded by areas that display a significant increase in
    calls, implying that most load was shifted onto nearby cell towers and was not lost.
	\label{fig:blackoutSpatialCallLoad}}
\end{figure}

Finally, there is also the concern that an emergency may be so severe that it interferes with the
operations of the mobile phone system itself.  Only the blackout caused any damage to the mobile
phone system, where some towers were temporarily disabled. (No calls appear to have been lost as
other towers picked up the slack.) See Fig.~\ref{fig:blackoutSpatialCallLoad}. Likewise, no towers
reached maximum capacity, preventing important calls from being routed.  Such effects may occur
during larger, more serious emergencies.


\subsection{Missing events}
It is possibile that newsworthy events may not be discoverable using mobile phones. Indeed, while
there are sixteen events documented in main text Table 1, there were a number of events discovered
in news reports that we could not identify in the data.  The majority of these were forest fires.
While they affected large regions and numbers of people, we could not find them with mobile phones.
A large wind storm and a gas main explosion were also not confirmed; both occurred late at night.
Two other events, a chemical leak causing an evacuation and a fire at a remote factory causing
noxious fumes were discovered in the dataset, but the affected populations were very small, so we
decided to discount them.

The absence of these events in the dataset provides important information about the strengths and
weaknesses of using mobile phones to study emergencies.  Since they rely on user activity, events
that occur late at night, when most people are asleep, may be difficult to study.  Likewise, events
that are severe but diffuse, providing a slight effect over a very broad area, may not be
distinguishable from the background of normal activity (although the earthquake is an exception to
this).  Events in remote locations with little cellular coverage will also be more difficult to
study than events in well-covered and well-populated regions.

Finally, since we are especially interested in studying how information propagates socially, we
avoided national events, such as televised sports matches or popular public holidays, as these make
distinguishing the different event populations $G_i$ unreliable.

\section{Source of call anomaly}\label{sec:sourcecallanomaly}
The anomalous call activity raises an important question: is the observed spike due to individuals
who normally do not use their phone in the event region and now suddenly choose to place calls, or
are those who normally use their phone in the respective timeframe prompted to call more frequently
than under normal circumstances? We determined in the event region (i) the relative change in the
average number of calls placed per user, $\Delta \rho / \normal{\rho}$, and (ii) the relative
increase in the number of individuals that use their phone in this period $\Delta N / \normal{N}$.
Figure~\ref{Mfig:combinedTimeSeries:changeNrho} shows that during the bombing we see a 36\% increase
in phone usage whereas the number of individuals that make a call increases by 232\%.  Other
emergencies show a similar pattern: the plane crash, earthquake and blackout show increases in
$\rho$ $\left(N\right)$ of $21\% \left(67.5\%\right)$, $1.36\% \left(17.4\%\right)$ and $4.97\%
\left(20.8\%\right)$, respectively. Taken together, these results indicate that the primary source
of the observed call anomaly is a sudden increase of calls by individuals who would normally not use
their phone during the emergency period, a behavioral change triggered by the witnessed event.

\section{Whom do people call}\label{sec:zF}
To see if affected users tend to call existing friends or contact strangers, we measured the
probability $P$ for a user in $G_0$ to make his first call between $t_\mathrm{start}$ and
$t_\mathrm{stop}$ to a friend, where `friends' are the set of individuals that have had phone
contact with the $G_0$ user during the previous full three months (not including the month of the
event).  Computing the mean and standard deviation of $P$ over normal time periods (all weekdays of
the month of the event, except the day of the event) allows us to quantify the relative change
during the event with 
\begin{equation}
	z_\mathrm{F} = \frac{P_\mathrm{event} - \avg{P_\mathrm{normal}} } { \sigma\left(P_\mathrm{normal}\right) }. 
\end{equation}
In all emergencies we observe an increase in the number of calls placed to friends (and a
corresponding decrease in calls to non-friends). Many non-emergencies show the opposite trend: users
are less likely to call a friend, although this change is seldom large.  See
Sec.~\ref{sec:systematicmechanisms} and Fig.~\lastFigNum{}d.

\section{Gender response during events}\label{sec:genderresponse}
\begin{figure*}\centering
	\includegraphics[width=0.7\textwidth]{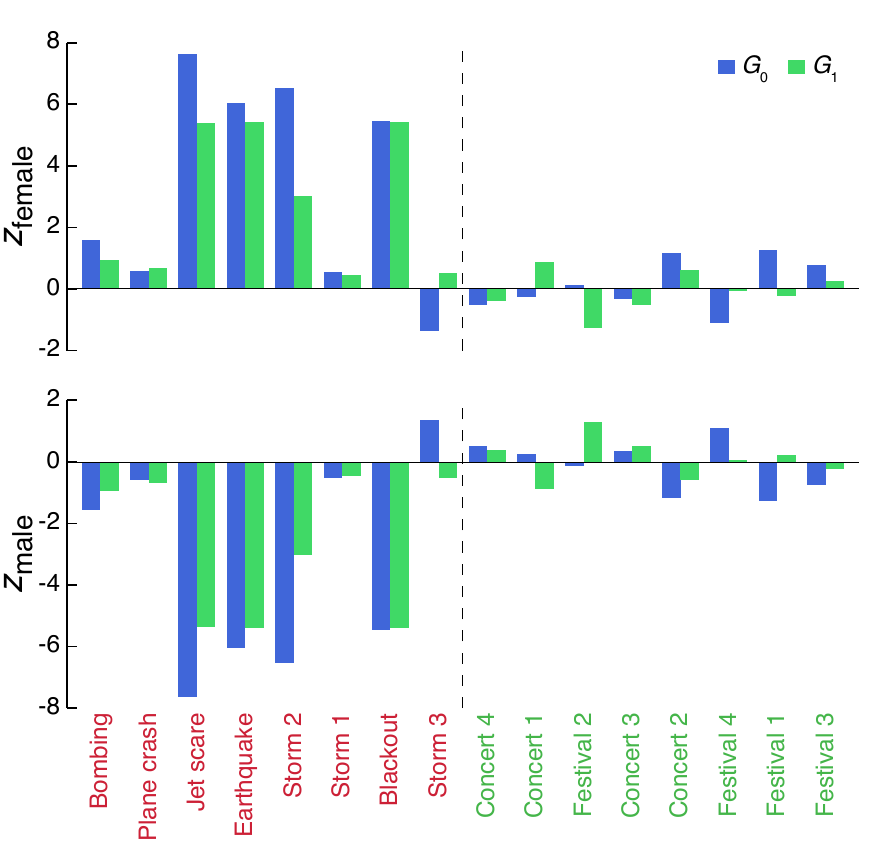}%
\caption[Gender response during emergency and non-emergency events]{\spacecmdFig{} \small  
    \textbf{Gender response during emergency and non-emergency events.} For each event we compute
    the significance $z_\mathrm{female}$ and $z_\mathrm{male}$ in the fraction of female and male
    users active during the event, compared with normal time periods, for both directly affected
    users ($G_0$) and those one step away ($G_1$).  We see that nearly all emergencies cause an
    increase in the fraction of affected female users; this increase is significant for half of the
    emergencies.  Non-emergencies do not result in deviations in the gender breakdown of affected
    populations.
	\label{fig:gender}%
}
\end{figure*}

To investigate how the population response to events depends on demographic factors, we used the
self-reported gender information, available for the majority ($\sim$88\%) of the users. 	For each
event we compute the significance $z_\mathrm{female}$ and $z_\mathrm{male}$ in the fraction of
female and male users active during the event, compared with normal time periods, as described in
\ref{sec:zF}, for both directly affected users ($G_0$) and those one step away ($G_1$).  As shown in
Fig.~\ref{fig:gender}, we see that nearly all emergencies cause an increase in the fraction of
affected female users; this increase is significant for half of the emergencies.  Non-emergencies do
not result in deviations in the gender breakdown of affected populations. These results hold for
both the directly affected users and their neighbors one step away.

\section{Voice versus text usage during events}
\begin{figure}[t]\centering{}
    \includegraphics[width=0.7\textwidth]{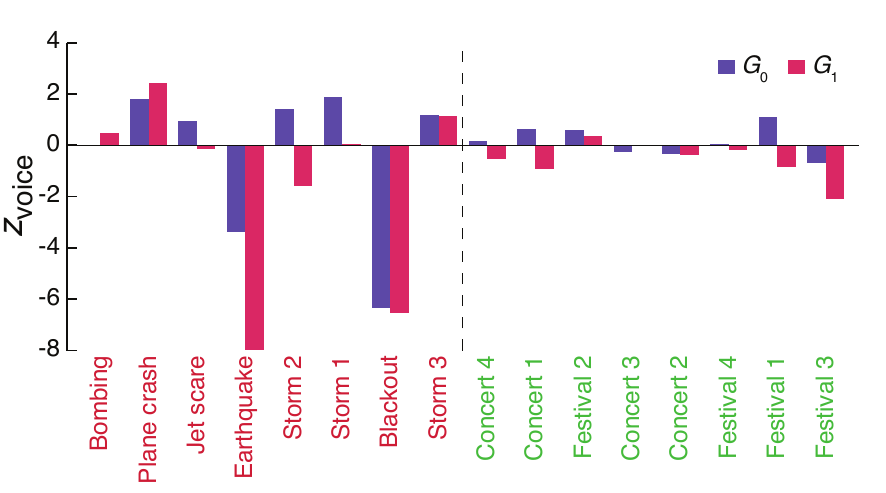}
\caption[Voice and text usage during emergencies]{\spacecmdFig{} \small
    \textbf{Voice and text usages during emergencies.}  Most events do not show a significant change
    in the fraction of voice calls compared to text messages.  The earthquake and blackout are
    exceptions, as is the plane crash and festival 3 ($G_1$ only).
    \label{fig:voicetext}}
\end{figure}

Similar to the quantities $z_\mathrm{F}$ and $z_\mathrm{female}$, we can assess whether users have
changed their means of communication due to an event.  To do so we compute the significance
$z_\mathrm{voice}$ of the fraction of voice calls compared to text messages during the event, for
both populations $G_0$ and $G_1$ (Fig.~\ref{fig:voicetext}).  The earthquake and blackout give
significantly increased text usage while the plane crash shows an increase in voice, while most
other events do not show a significant change.  Since the earthquake and blackout were both
relatively minor (low danger) events, this result implies that a spike in primarily text messaging
activity may indicate that the event is a low-threat/non-critical emergency.

\section{Systematic response mechanisms during emergencies}\label{sec:systematicmechanisms}

As mentioned in the main text, to summarize our current understanding of these events, we computed
temporal, spatial, and social properties for each anomaly, plotted in main text Fig.~\lastFigNum{}.
Temporally, we study the midpoint fraction:
\begin{equation}
 \fmid=\left(t_\mathrm{mid}-t_\mathrm{start}\right)/\left(t_\mathrm{stop}-t_\mathrm{start}\right),
\end{equation}
 the fraction of time required for half of the anomalous call activity to occur, where
 $t_\mathrm{mid}$ satisfies:
 \begin{equation}
 	\int_{t_\mathrm{start}}^{t_\mathrm{mid}}\big( V_\mathrm{event}(t) - V_\mathrm{normal}(t)\big) \, dt
 	= \frac{1}{2}\int_{t_\mathrm{start}}^{t_\mathrm{stop}}\big( V_\mathrm{event}(t) - V_\mathrm{normal}(t)\big) \, dt.
 \end{equation}
The midpoint fraction is more robust to noisy and non-sharply peaked time series, where estimating
$t_\mathrm{peak}$ is difficult, than the peak fraction. Spatially, we use the anomaly's $\rc$.
Socially, we compute two quantities: the relative size of the cascade $R$ (the number of people in
the event cascade divided by the number generated under normal periods; see
Sec.~\ref{sec:calcSocProp:controls} and Fig.~\ref{fig:compareRs}) and $z_\mathrm{F}$, the
significance in the probability of calling a friend compared with a non-friend (see
Sec.~\ref{sec:zF}).

Figure \lastFigNum{} shows a distinct separation in these measures between emergencies and
non-emergencies, indicating that there are universal response patterns underlying societal dynamics
independent of the particular event details.

\section{Calculating social propagation}\label{sec:calcSocProp}
In this section we detail the procedure for extracting the contact network between users after an
event (Sec.~\ref{sec:calcSocProp:contactNet}) and how to control for various factors to demonstrate
whether or not the contact network or its information cascade is anomalous due to the event
(Sec.~\ref{sec:calcSocProp:controls}).

\subsection{Constructing the time-dependent contact network}\label{sec:calcSocProp:contactNet}

\begin{figure}[tp]\centering
	\subfloat{\label{fig:extractContactNet:callLogs}}
	\subfloat{\label{fig:extractContactNet:timeEvolve}}
	\includegraphics[width=1\textwidth]{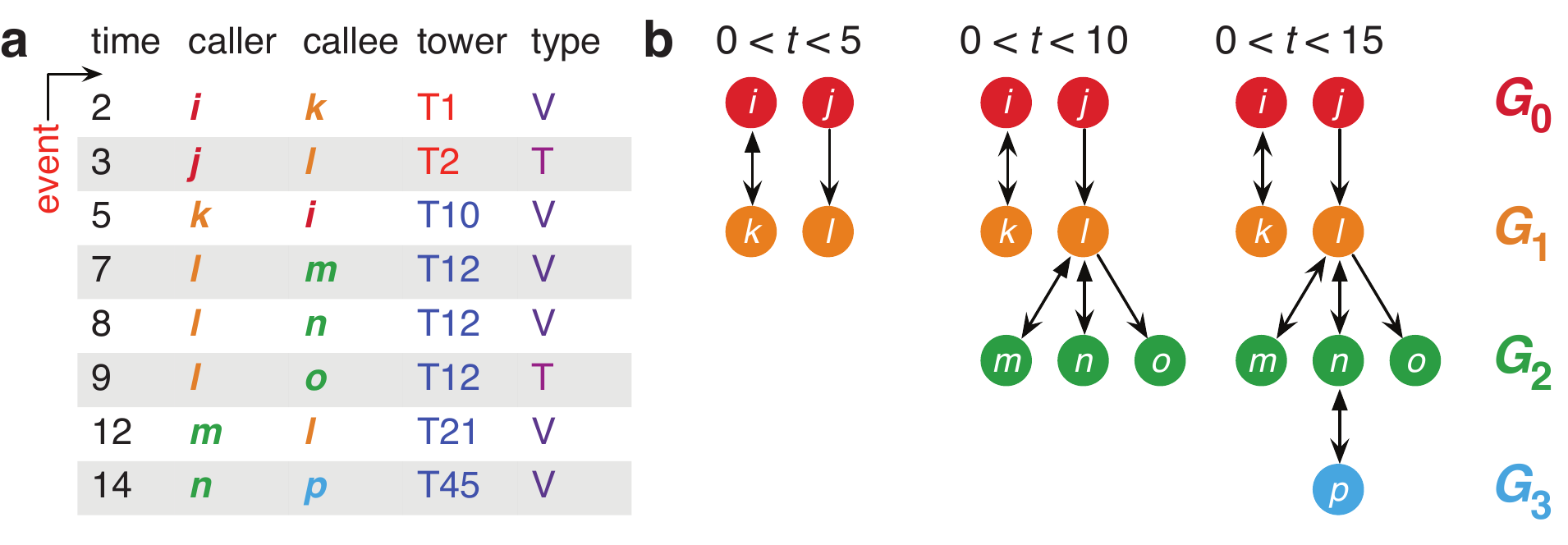}
	\caption[Extracting the time-dependent contact network from call data]{\spacecmdFig{} \small
	\textbf{Extracting the time-dependent contact network from call data}.
    \sciLet{\subref*{fig:extractContactNet:callLogs}} A cartoon example of the dataset's call
    records, representing eight calls during 15 minutes following an event.  The region of the event
    contains two towers, T1 and T2.  Call type is (T) for text message and (V) for voice call.
    \sciLet{\subref*{fig:extractContactNet:timeEvolve}} Three instances of the evolving contact
    network, extracted from the example call log shown in \textbf{a}.  Two users made calls from the
    region during the first five timesteps, initiating a cascade.
	\label{fig:extractContactNet}}
\end{figure}

We use the following process to generate the contact network between users due to an event. We
follow all messages in order of occurrence during the event's time interval
$\left[t_\mathrm{start},t_\mathrm{stop}\right]$.  While we do not know the content of the messages,
we assume that any related messages do transmit information pertaining to the event. A user $u$ who
does not know about the event becomes ``infected'' with knowledge due to communication at some time
$t \in \left[t_\mathrm{start},t_\mathrm{stop}\right]$ if (1) $u$ initiated communication from a
tower within the event region or (2) $u$ communicated with a user that was already infected.  We
place $u$ in the set $G_0$ if $u$ communicated from the event region, otherwise we place $u$ into
$G_{i+1}$ where the infected user transmitting knowledge to $u$ was in $G_i$.  Following these calls
and generating the $G_i$ with this procedure then forms the contact network.  Note that there are
two types of communications in the dataset, voice and text. We assume that voice is bidirectional
whereas text messages are not (a user who sends a text message to someone with knowledge of the
event will learn nothing from that particular communication).  An illustration of this process is
depicted in Fig.~\ref{fig:extractContactNet}.

\begin{figure}%
\begin{minipage}{5.9cm}%
	\centering%
	\caption[Distribution of shortest paths within bombing's contact network]{\spacecmdFig{} \small
    \textbf{Distributions of shortest paths found within the bombing's contact network.}  One can
    compute shortest paths emanating from all users (\textbf{top}) or only users in $G_0$
    (\textbf{bottom}) and for paths within the entire network (\textbf{left}) or only the network's
    giant component (\textbf{right}).
	\label{fig:bombing_shortest_paths}}%
\end{minipage}\hfill%
\begin{minipage}[]{10cm}\centering%
	\includegraphics[width=1\textwidth]{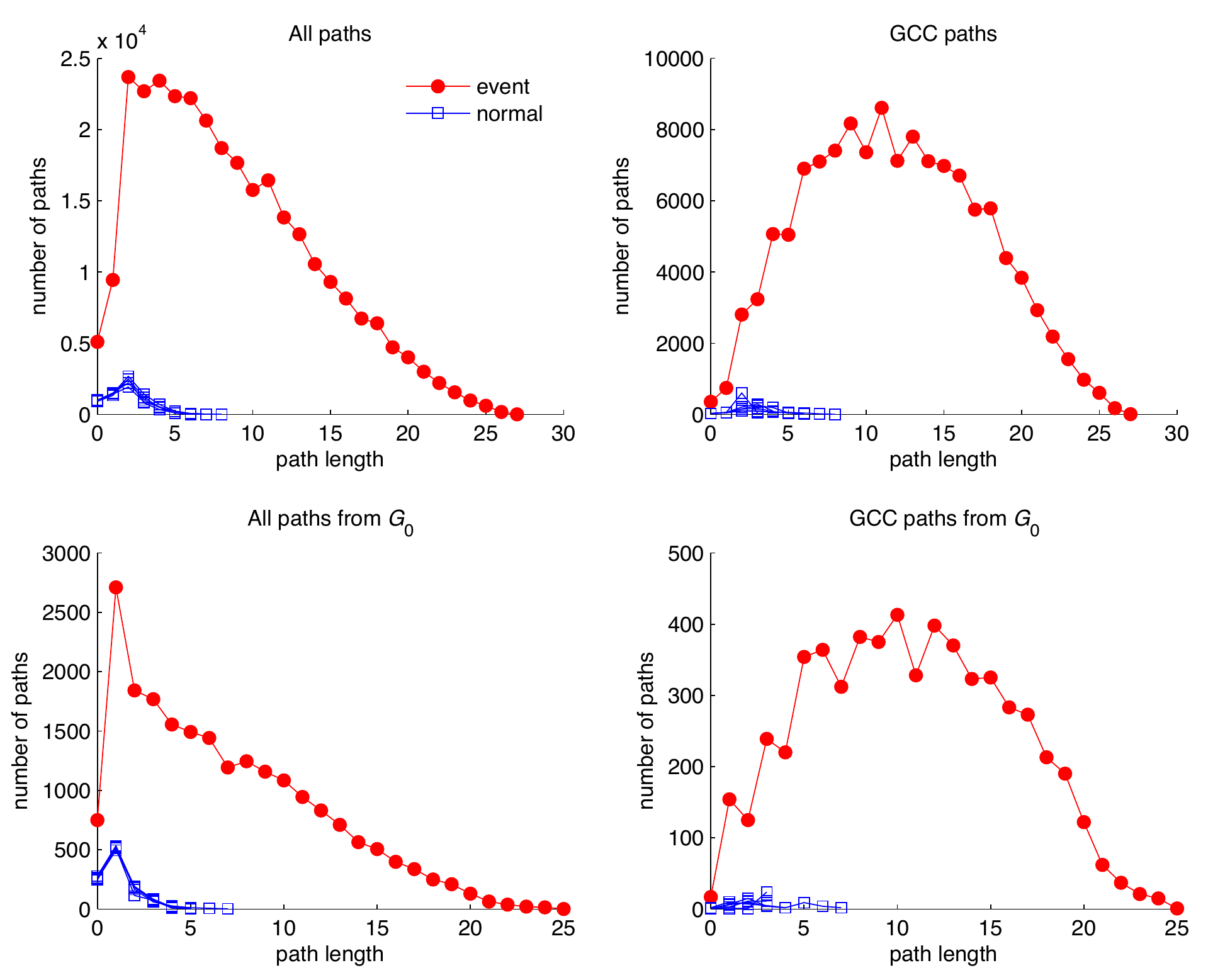}%
\end{minipage}
\end{figure}

The contact network itself can be studied using a number of network science tools.  One way to
analyze the size and scale of this network is through the distribution of shortest (or geodesic)
path lengths~\cite{newmanSIAMreview}.  In Fig.~\ref{Mfig:incSocialDist:bombPaths} we presented the
distribution of paths emanating from $G_0$ users within the giant connected component (GCC) of the
bombing's network.  One can also analyze the distribution for all users, not just $G_0$, and for all
components of the network.  These possibilities are shown in Fig.~\ref{fig:bombing_shortest_paths}.

Finally, the cascade of information through a contact network is a non-local process which may be
highly effected by sampling/percolation~\cite{newmanSIAMreview}.  Indeed, the mobile phone dataset
contains only users of a single phone company, and a number of propagation paths may be missing.
However, the dataset actually contains all users who make or receive calls to users within the
company, even those outside the company.  This means we have all cascade paths of one or two steps
that begin and end with in-company users (regardless of whether they travel through a company user
or not) and that those paths are the actual shortest paths, providing an effective lower bound on
the cascade.  In other words, if we can demonstrate the existence of a cascade over users
$\left\{G_0,G_1,G_2\right\}$, then the actual cascade containing those users can only be larger.

\subsection{Controlling for social propagation}\label{sec:calcSocProp:controls}

A contact network between users can always be constructed, even when no event takes place.  There
may appear to be propagating cascades as well, since there are temporal correlations between users
receiving and then placing calls.  This must be properly controlled for.

Suppose we are studying an event and have identified its affected users $G_0$, those users that made
calls during some time window $(t,t+\Delta t)$. We expect that $G_0$ users will call other users
($G_1$) outside the event, $G_1$ users will call $G_2$ users, etc., generating a cascade
$\{G_0,G_1,\ldots\}$. Tracking calls starting from some $G$ will always generate such groups, even
during normal periods. Then the question is, do the users in $G_i, i>0$ show increased activity
during an emergency or other event?  If so, that is evidence of the social propagation of
situational awareness.

\begin{figure}[t]%
	 \subfloat{\label{fig:understanding_biases_SI:CallsFullCity}}          
	 \subfloat{\label{fig:understanding_biases_SI:CallsRandPop}}           
	 \subfloat{\label{fig:understanding_biases_SI:CallsRandPopsDiffSizes}} 
	\centering{}%
	\includegraphics[width=1\textwidth]{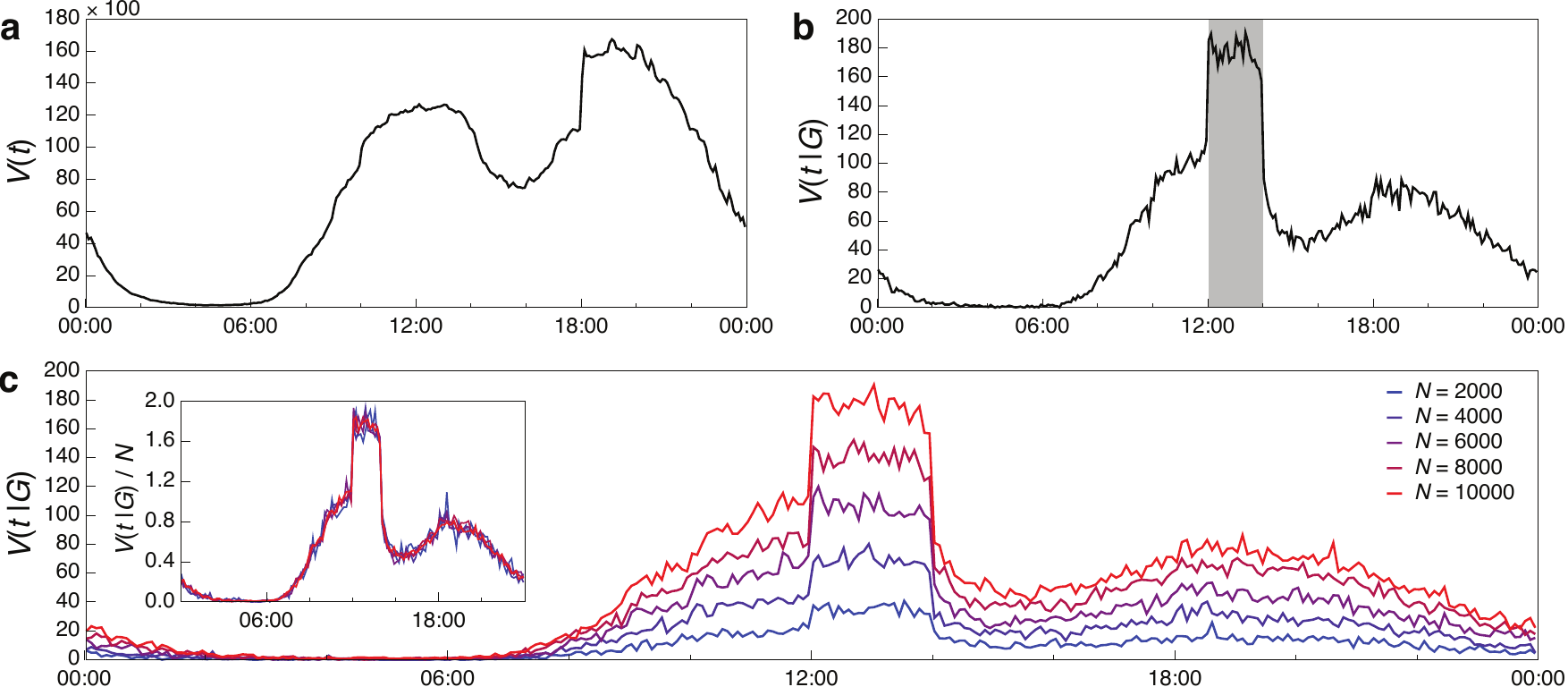}%
\caption[Understanding selection bias]{\spacecmdFig{} \small
	\textbf{Understanding selection bias.}
    \sciLet{\subref*{fig:understanding_biases_SI:CallsFullCity}} The call volume $V(t)$ of a major
    city during an ordinary 24-hour period.
    \sciLet{\subref*{fig:understanding_biases_SI:CallsRandPop}} The call volume $V(t|G)$ of $N=10^4$
    randomly selected users from \textbf{\subref*{fig:understanding_biases_SI:CallsFullCity}} who
    all placed one or more calls between 12:00 and 14:00 (highlighted).  The `bias' of this
    conditional time series is clear.
    \sciLet{\subref*{fig:understanding_biases_SI:CallsRandPopsDiffSizes}} The same as
    \textbf{\subref*{fig:understanding_biases_SI:CallsRandPop}} for different values of $N$.
    Rescaling by the population size  (inset) indicates that the relative scale of the bias of
    $V(t|G)$ during the time window is independent of the population size.
	\label{fig:understanding_biases_SI}}
\end{figure}

To answer this question we need to consider several points:
\begin{itemize}
    \item Suppose that the total call activity for a region is $V(t)$
        (Fig.~\ref{fig:understanding_biases_SI:CallsFullCity}).  Now select a group of users
        $G_{\Delta t}$ that each place at least one call during a small time window $\Delta t$.
        Tracking only their call activity generates the conditional time series $V(t|G_{\Delta t})$.
        This time series has a ``selection bias'' that creates the appearance of a large increase in
        activity during the time window since all the users must place calls then
        (Fig.~\ref{fig:understanding_biases_SI:CallsRandPop}).  This must be accounted for when
        studying selected users during an event.
	
    \item When tracking $V(t|G)$ for a group of $N=|G|$ users, the overall level of activity will
        depend on $N$ (Fig.~\ref{fig:understanding_biases_SI:CallsRandPopsDiffSizes}).  A rescaling
        is necessary when comparing the activity levels of different size groups.
	
    \item The selection bias will also depend on the length of the time window $\Delta t$.  If
        $\Delta t=24$ hours had been used in Fig.~\ref{fig:understanding_biases_SI}, no bias would
        be evident.  All events are compared to time periods with the same $\Delta t$, so this
        effect is automatically controlled for.
\end{itemize}

Having considered these aspects, we turn our attention to the problem of calculating the cascade
itself.  For the event period, tracking the outgoing calls of $\{G_0,G_1,\ldots\}$ is
straightforward.  To determine how unusual this activity is, we need controls for comparison.  There
are several possibilities:
\begin{description}
    \item[Control 1] One option is scanning the event region during the same time of the week,
        collecting a control population of users, and then following their cascade.  However, this
        does not account for changes in the composition or number of users in the event region (some
        studied events were quite remote and typically contained very few users). 
    \item[Control 2] Another possibility is to simply follow the activity of the \emph{same} users
        $G_i$ from the event's cascade during normal time periods.  This choice is keeps the
        population unchanged but it does not account for changes in who is being called; $G_0$ users
        may have chosen to call very different people during an emergency.  Further, it does not
        account for the selection bias that is present during the event but not during the normal
        periods, which may exaggerate the change in call activity.
    \item[Control 3] Finally, one can study new cascades generated by the \emph{same} event users
        $G_0$ during normal periods, creating a different cascade $\{G_0,g_1,g_2,\ldots\}$ for
        \textbf{each} normal period.  The activities of each $G_i$ can then be compared to those of
        the corresponding $g_i$'s. This directly tests the effect that the initiating population
        $G_0$ has on the cascade, by studying those cascades the population would normally induce,
        and accounts for selection bias since this bias is present during the event and the normal
        periods.
\end{description}

We have chosen to use Control 3.  Note that $G_i$ will typically be larger than the normal $g_i$'s
and that $G_i$ users may be more active than those in $g_i$, so $V(t|g_i)$ must be \textbf{rescaled}
when being compared to the event's $V(t|G_i)$.  To do this, we multiply $V(t|g_i)$ by a constant
scaling factor $a_i$,
\begin{equation}
	a_i = \left. \int_{\delta t} V(\tau|G_i)\,d\tau \middle/ \int_{\delta t} V(\tau|g_i)\,d\tau \right.,
	\label{eqn:rescalingfactors}
\end{equation}
where both integrals run over the same ``calibration interval'' ${\delta t}$ and $\tau=0$ is the
start of the selection window.  For most events we integrate over a 24-hour period two days before
the window, $\delta t = (-48,-24)$. If the event is on a weekday, we ensure the calibration interval
is not a weekend and vice versa.  This factor $a_i$ was chosen such that the total number of calls
during normal time periods for $V(t|G_i)$ is approximately equal to $a_i V(t|g_i)$, equalizing the
smaller time series and removing bias due to $|G_i| \neq |g_i|$.

\begin{figure}[]%
\begin{minipage}[]{8cm}\centering%
	\subfloat{\label{fig:compareRs:lessControls}}
	\subfloat{\label{fig:compareRs:moreControls}}
	\includegraphics[]{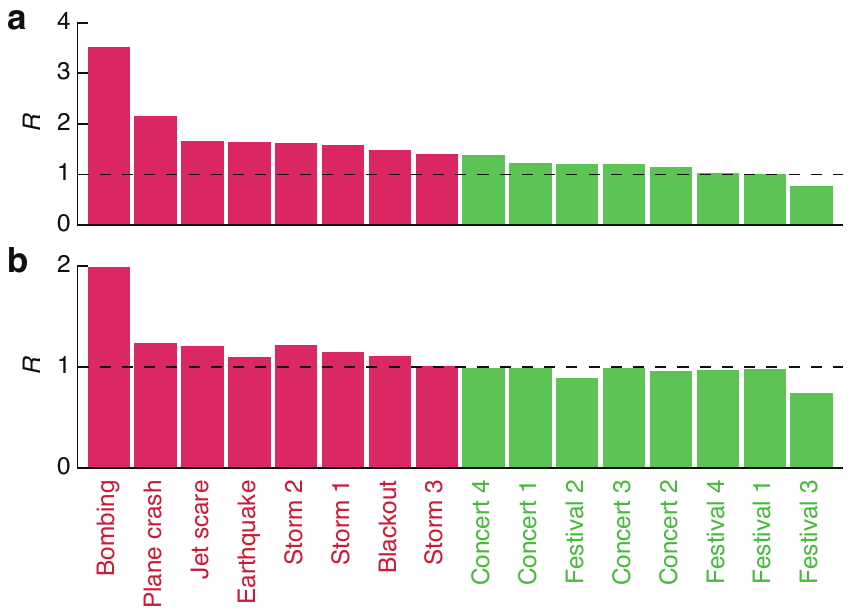}
\end{minipage}\hfill%
\begin{minipage}{7.25cm}%
	\centering%
	\caption[Controlling factors that affect the relative cascade size]{\spacecmdFig{} \small%
	\textbf{Controlling factors that affect the relative cascade size.}
	\sciLet{\subref*{fig:compareRs:lessControls}}
    Relative cascade size $R=\event{N} / \normal{N}$, where $\event{N}=\sum_i |G_i^e|$ and
    $\normal{N} = |G_0^e| + \langle\sum_{i>0} \left|g_i^n\right| \rangle_n $ (averaged over normal
    events $n$).
	\sciLet{\subref*{fig:compareRs:moreControls}}
    Relative cascade size using $R = \langle \sum_i |g_i^{e_n}| / \sum_j |g_j^n| \rangle_n$,
    controlling for the lower activity level during normal period $n$ with $g_0^{\mathrm{e}_n}
    \subset G_0^\mathrm{e}$.  With this control, all non-emergencies have $R<1$ and all emergencies
    have $R\gtrsim 1$ (Storm 3 has $R=1.0044$).
	\label{fig:compareRs}
	}
\end{minipage}%
\end{figure}

Control 3 and Eq.~\ref{eqn:rescalingfactors} allow one to compare disparate populations' call
activities, but there are two subtle yet important details to consider when comparing event and
normal contact networks:
\begin{enumerate}

\item The event contact network under Control 3 consists of users
    $\left\{G_0^\mathrm{e},G_1^\mathrm{e},G_2^\mathrm{e},\ldots\right\}$ while the contact network
    for normal period $n$ contains users $\left\{G_0^\mathrm{e},g_1^n,g_2^n,\ldots\right\}$.  As
    shown in Fig.~\ref{Mfig:combinedTimeSeries:changeNrho}, many members of $G_0^\mathrm{e}$ will be
    silent during each period $n$. This means that the cascade during period $n$ is
    \textbf{actually} generated by a smaller set of users $g_0^{\mathrm{e}_n} \subset
    G_0^\mathrm{e}$, defined as those users in $G_0^\mathrm{e}$ who also make calls during period
    $n$.  The rescaling factors $a_i$ account for this when comparing activity levels $V(t|G)$ but
    this may give an unfair advantage to the growth of the event contact network, as quantified by
    the relative cascade size $R$, when compared to normal periods. For example, an event may result
    in $R>1$ simply because the event period's ``seed'' population is larger than that of the normal
    period.  See Fig.~\ref{fig:compareRs:lessControls}.

    To control for this effect, we compute $G_0^\mathrm{e}$ for the event and use it to generate
    $g_0^{\mathrm{e}_n}$ for each period $n$.  We then return to the original event period and track
    a new event cascade for each $g_0^{\mathrm{e}_n}$.  In other words, we consider only the smaller
    cascade $\left\{g_0^{\mathrm{e}_n},g_1^{\mathrm{e}_n},\ldots\right\}$ formed by following only
    users who are normally active during that time of day.  This means there are now different event
    cascades corresponding to each normal period. See Fig.~\ref{fig:compareRs:moreControls}.

\item There is a tendency for non-emergencies to occur later in the day than the emergency events.
    Most concerts are held at night, for example.  Users are generally more active during this time
    of day; many users have plans with free nightly minutes.  This factor is not an issue for
    activity levels $V(t)$ since these are always compared during the same time of day.  However,
    this may give an additional advantage when e.g., comparing $R$ for an event that occurs early in
    the morning to an event that occurs during peak call activity.  The latter cascade occurs when
    users are generally more active, increasing the number of potential propagation vectors and
    possibly increasing the rate at which new members are added to the $G_i$'s.

    A simple way to control for this is the following.  First, compute the Global Activity Level
    (GAL), the total number of phone calls in the entire dataset during each event's time window.
    Then, pick the event that contains the lowest GAL and shorten all the time windows (decreasing
    $t_\mathrm{stop}$ only) for the other events such that they have the same minimum GAL.  Finally,
    generate the contact network during these reduced windows only.  Doing this guarantees that
    every event has the same number of possible communications to propagate along.  This process was
    used for all computations of $R$.

\end{enumerate}

With this machinery in hand, we can now successfully demonstrate that there are anomalously large
cascades for many emergencies, especially the bombing.  Not only is $R>1$ (Figs.~\lastFigNum{}c and
\ref{fig:compareRs}) but the contact populations show anomalous increases in call activity, even
after all controls are in place (Figs.~\ref{Mfig:incSocialDist:dVforSomeGi},
\ref{Mfig:incSocialDist:dVintegratedVsGi}, \ref{fig:socProp_4mainEmergs_SI}, and
\ref{fig:socProp_4otherEmergs_SI}).  All non-emergencies generate ordinarily-sized contact networks
(Figs.~\lastFigNum{}c and \ref{fig:compareRs}) and normal activity levels
(Figs.~\ref{Mfig:incSocialDist:dVforSomeGi}, \ref{Mfig:incSocialDist:dVintegratedVsGi},
\ref{fig:socProp_4concerts_SI}, and \ref{fig:socProp_4festivals_SI}).

\begin{table}[htbp]\centering\small
\begin{tabular}{llllll}
\toprule 
                                      &    & Event 			   		& GAL        & $t_\mathrm{stop}$ (min) & $t_\mathrm{stop}'$ (min) \\ %
\midrule                                                                                                                         
\vrtLbl{\bf\color{red}Emergencies}    & 1  & \textit{Bombing}	   	& 2,327,592  & 120                     & 65                       \\ %
                                      & 2  & \textit{Plane crash} 	& 1,497,402  & 120                     & 99                       \\ %
                                      & 3  & \textit{Earthquake}  	& 1,370,268  & 60                      & 60                       \\ %
                                      & 4  & \textit{Blackout}	   	& 4,536,253  & 180                     & 51                       \\ %
                                      & 5  & Jet scare  		   	& 1,320,363  & 80                      & 77                       \\ %
                                      & 6  & Storm 1			   	& 3,701,872  & 135                     & 48                       \\ %
                                      & 7  & Storm 2			   	& 1,266,468  & 115                     & 115                      \\ %
                                      & 8  & Storm 3			   	& 2,654,849  & 120                     & 51                       \\ %
\midrule                                                                                                                              
\vrtLbl{\bf\color{green}Non-emergencies} & 9  & \textit{Concert 1}	   	& 8,136,294  & 360                     & 46                       \\ %
                                         & 10 & Concert 2			   	& 6,512,493  & 240                     & 45                       \\ %
                                         & 11 & Concert 3			   	& 10,384,830 & 540                     & 49                       \\ %
                                         & 12 & Concert 4			   	& 6,473,695  & 690                     & 42                       \\ %
                                         & 13 & \textit{Festival 1}  	& 13,782,768 & 960                     & 74                       \\ %
                                         & 14 & Festival 2		      	& 6,095,958  & 1,200                   & 42                       \\ %
                                         & 15 & Festival 3		      	& 19,504,204 & 1,200                   & 90                       \\ %
                                         & 16 & Festival 4		      	& 10,386,001 & 730                     & 89                       \\ %
\bottomrule
\end{tabular}
	\caption[Global Activity Level and stop times for all sixteen events]{\spacecmdFig{} \small%
    \textbf{Global Activity Level (GAL) and stop times for all sixteen events.} 
    show the original $t_\mathrm{stop}$ and modified $t_\mathrm{stop}'$ that equalizes all events'
    GAL.  Storm 2 had the smallest GAL over the corpus.}
\end{table}

\section{Results on the event corpus}\label{sec:results_corpus}

\begin{figure}[t]\centering{}
	\includegraphics[width=0.85\textwidth,trim=0 8 0 0,clip=true]{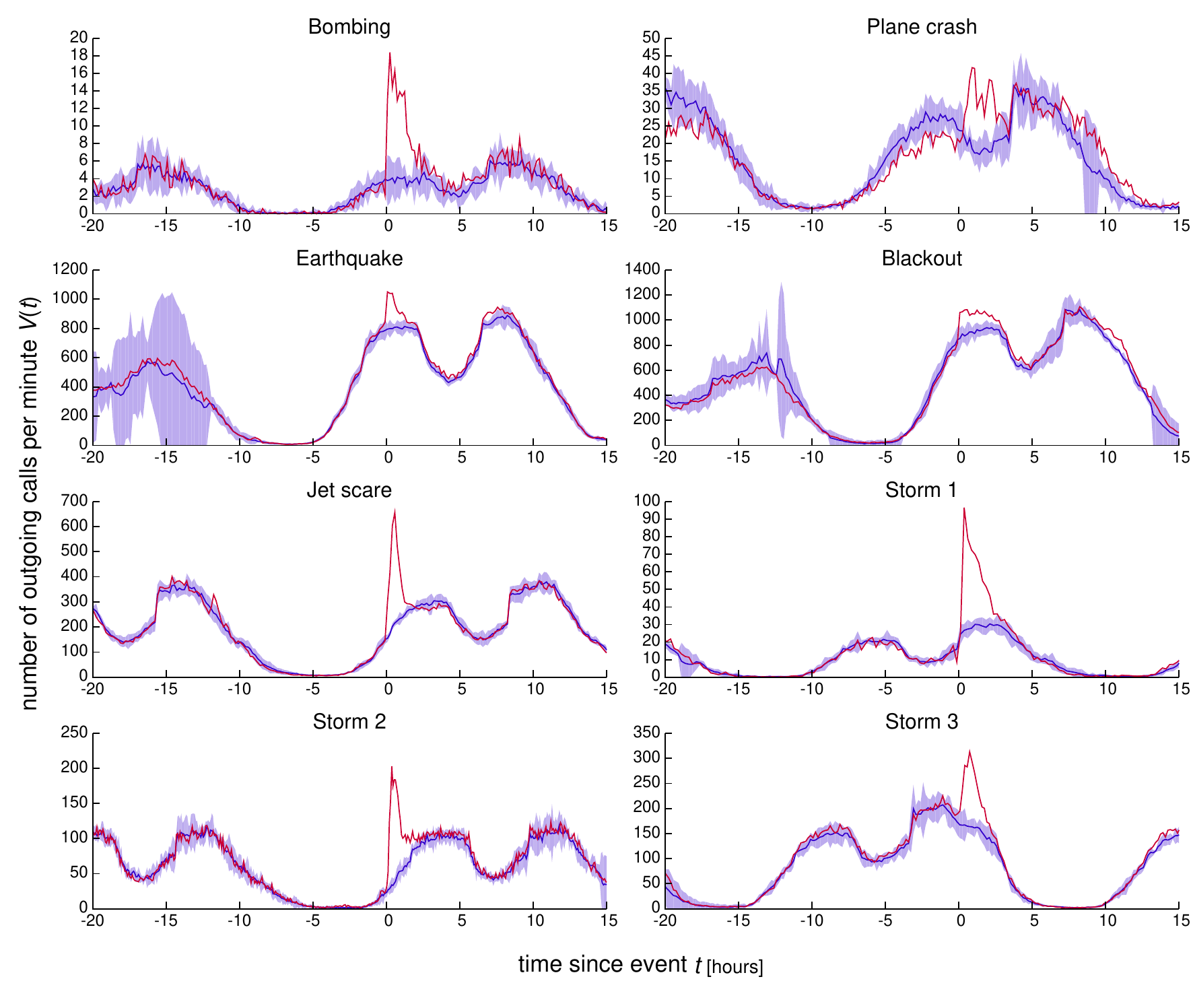}
	\caption[Regional call activity for the eight emergencies]{\spacecmdFig{} \small
    Regional call activity for the eight emergencies analyzed.  The first four are also shown in
    main text Fig.~\ref{Mfig:combinedTimeSeries:RawTimeSeries}.  Shaded regions indicate $\pm 2$
    standard deviations.\label{fig:timeseries_events}}
\end{figure}

\begin{figure}[t]\centering{}
	\includegraphics[width=0.85\textwidth,trim=0 8 0 0,clip=true]{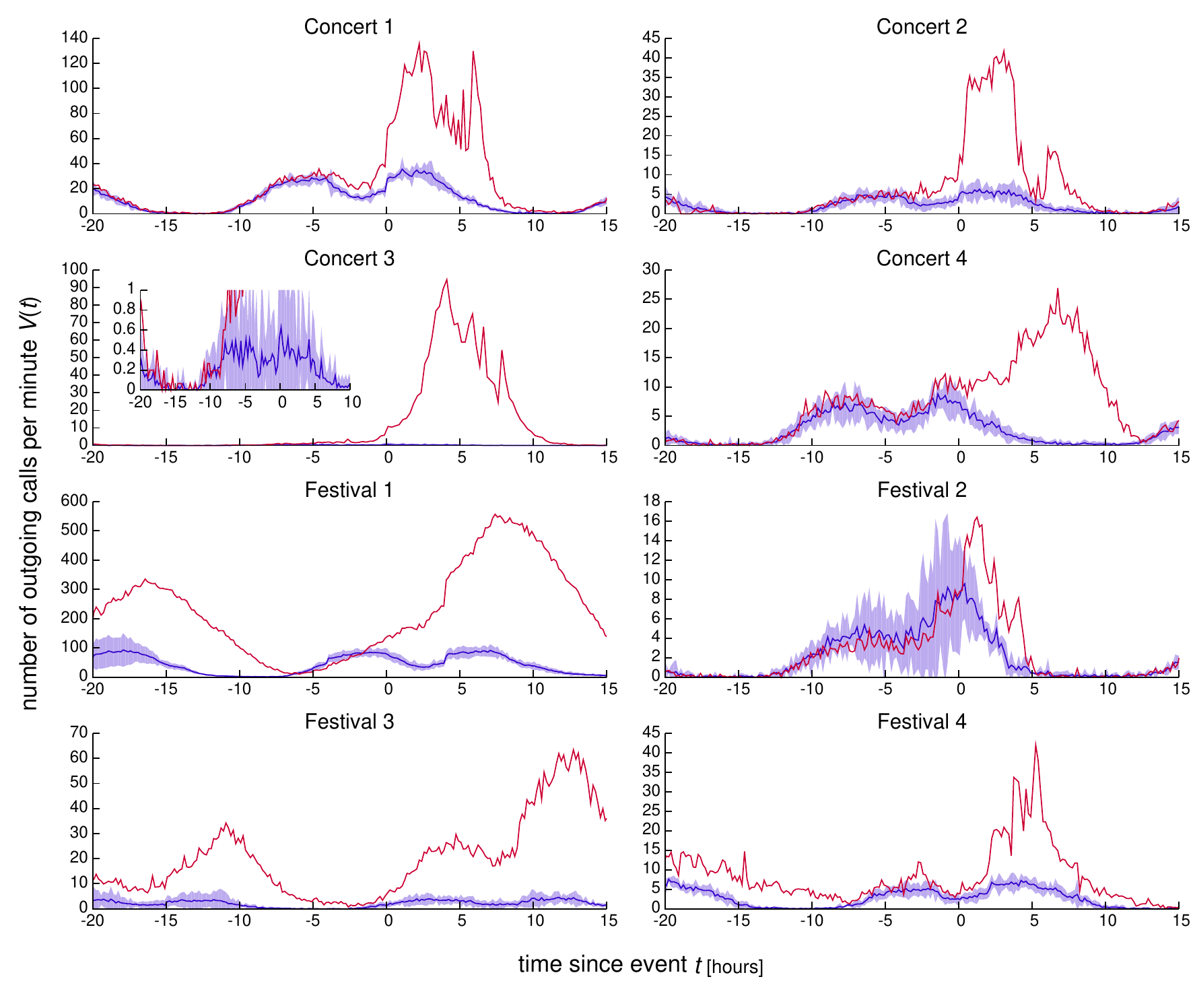}
	\caption[Regional call activity for the eight controls]{\spacecmdFig{} \small 
    The same as Fig.~\ref{fig:timeseries_events} for the eight non-emergencies.  Concert 3 takes
    place at an otherwise unpopulated location and the normal activity is not visible on a scale
    showing the event activity.
	\label{fig:timeseries_controls}
	}
\end{figure}

\begin{figure}[t]\centering{}
	\includegraphics[width=1\textwidth]{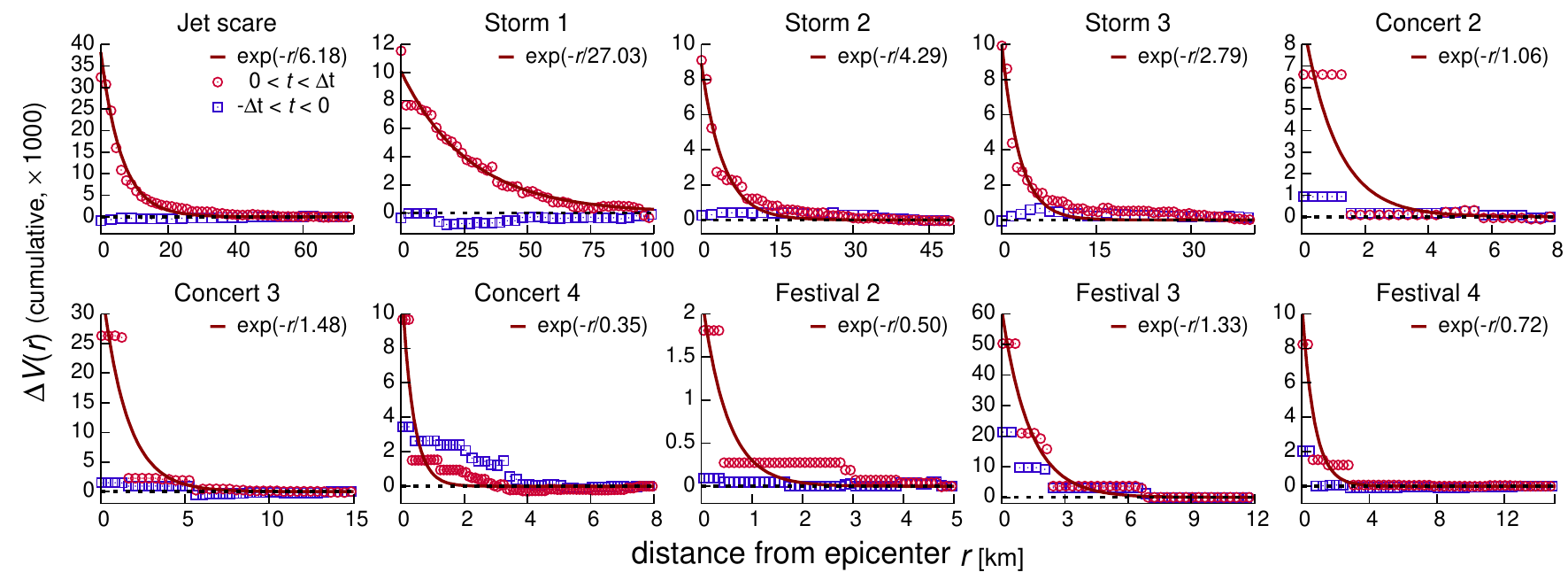}
	\caption[Spatial call activity for remaining events]{\spacecmdFig{} \small
    Spatial call activity for the ten events not shown in main text
    Fig.~\ref{Mfig:spatialProps:dVintegratedVsR}.\label{fig:spatialPropsSI}}
\end{figure}

\begin{figure}[t]\centering{}
	\includegraphics[width=1\textwidth]{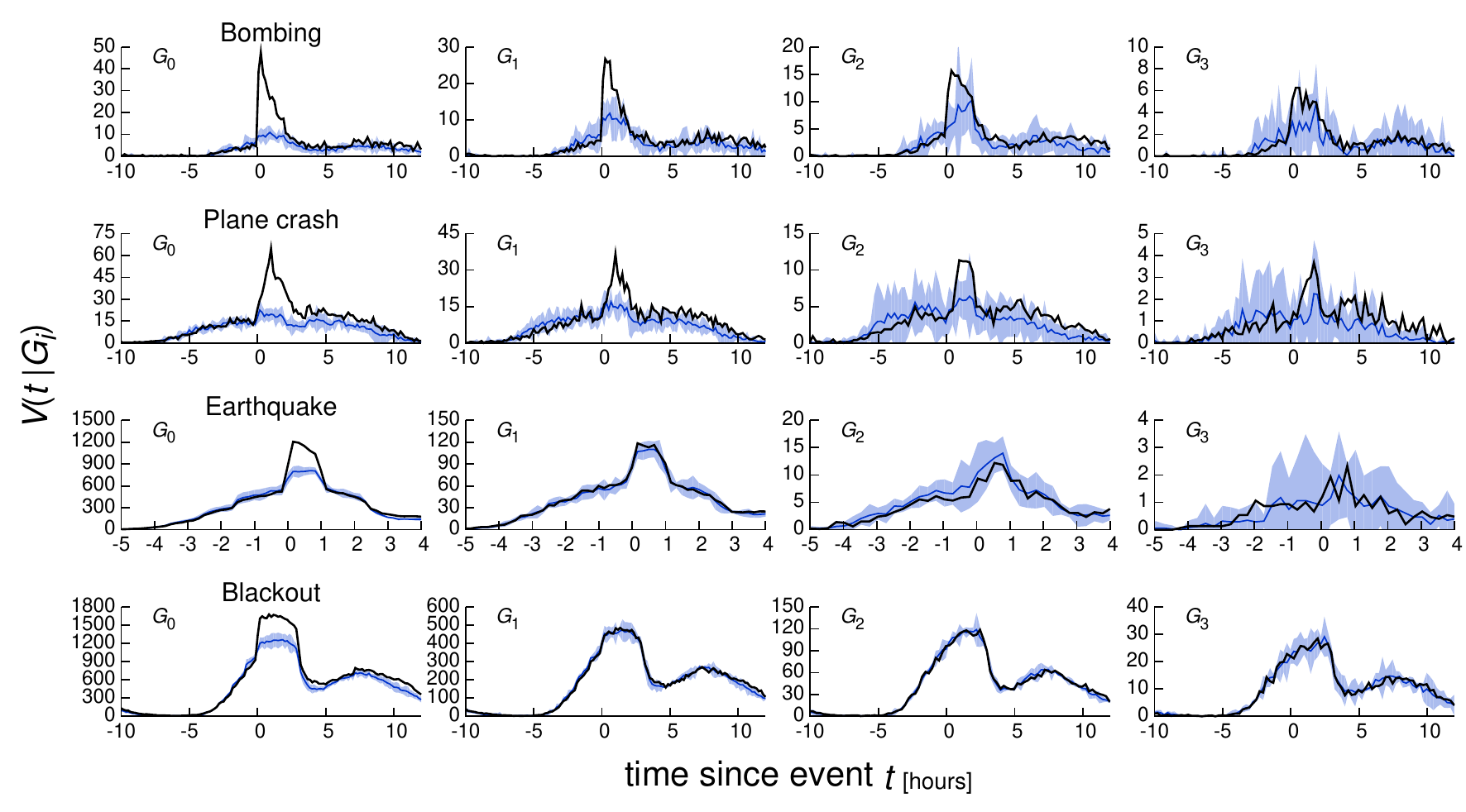}
	\caption[Social propagation for emergencies 1--4]{\spacecmdFig{} \small
    Social propagation for the four main text emergencies.  Shown are the activity patterns
    (conditional time series) for $G_0$ through $G_3$ during the event (black curve) and normally
    (shaded regions indicate $\pm 2$ s.d.).  Normal activity levels were rescaled to account for
    population and selection bias (see Sec.~\ref{sec:calcSocProp}) The bombing and plane crash show
    increased activities for multiple $G_i$ while the earthquake and blackout do not.
    \label{fig:socProp_4mainEmergs_SI}}
\end{figure}

\begin{figure}[t]\centering{}
	\includegraphics[width=1\textwidth]{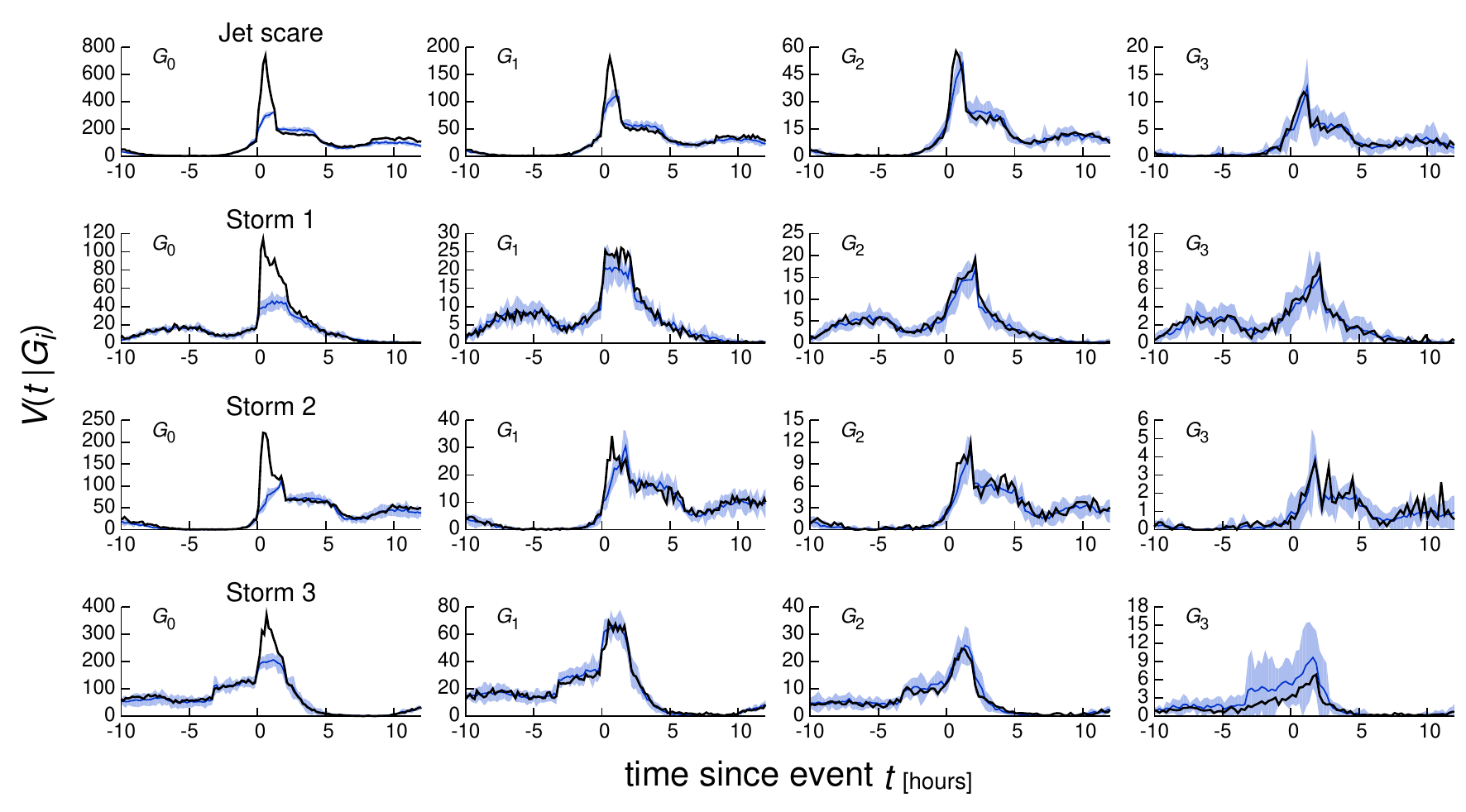}
	\caption[Social propagation for emergencies 5--8]{\spacecmdFig{} \small
    Same as Fig.~\ref{fig:socProp_4mainEmergs_SI} for the remaining emergency
    events.\label{fig:socProp_4otherEmergs_SI}}
\end{figure}

\begin{figure}[t]\centering{}
	\includegraphics[width=1\textwidth]{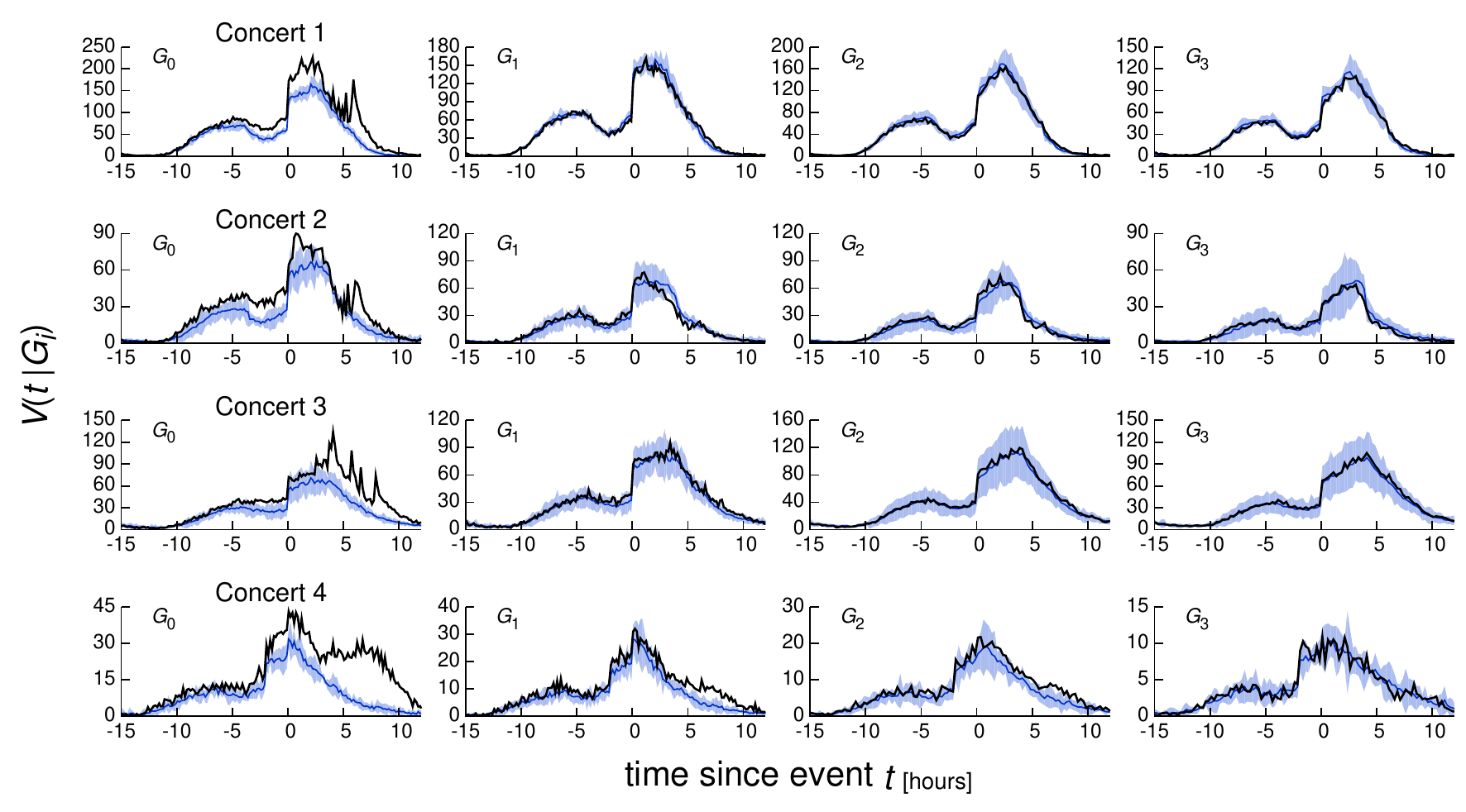}
	\caption[Social propagation for concerts]{\spacecmdFig{} \small
    Same as Fig.~\ref{fig:socProp_4mainEmergs_SI} for the concerts.  All concerts show extra
    activity only for $G_0$ except Concert 4, which shows a small increase in activity for $G_1$
    several hours after the concert started. \label{fig:socProp_4concerts_SI} }
\end{figure}

\begin{figure}[t]\centering{}
	\includegraphics[width=1\textwidth]{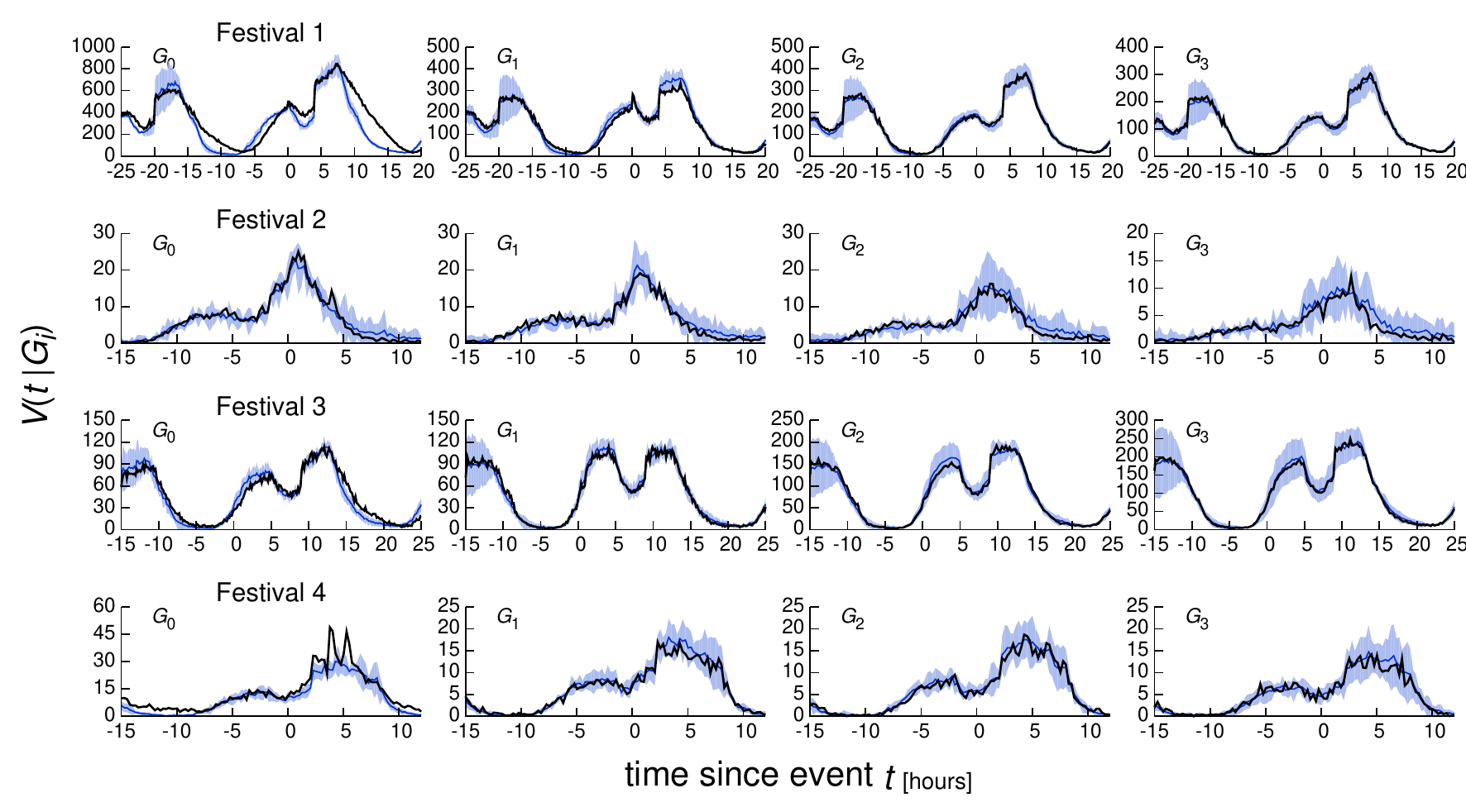}
	\caption[Social propagation for festivals]{\spacecmdFig{} \small
    Same as Fig.~\ref{fig:socProp_4mainEmergs_SI} for the festivals. Interestingly, Festival 2 shows
    no extra activity, even for $G_0$, indicating that the call anomaly for those events was caused
    only by a greater-than-expected number of users all making an expected number of calls.
    \label{fig:socProp_4festivals_SI}}
\end{figure}

Sixteen events were identified for this work (see main text Table 1), but six events were focused
upon in the main text.  Here we report the results for all events.  In
Figs.~\ref{fig:timeseries_events} and \ref{fig:timeseries_controls} we provide the call activities
$V(t)$ for all sixteen events used in this study (compare to Fig.~1). In
Fig.~\ref{fig:spatialPropsSI} we show $\Delta V(r)$ for the ten events not shown in main text
Fig.~\ref{Mfig:spatialProps:dVintegratedVsR}.  Finally in Figs.~\ref{fig:socProp_4mainEmergs_SI},
\ref{fig:socProp_4otherEmergs_SI}, \ref{fig:socProp_4concerts_SI}, and
\ref{fig:socProp_4festivals_SI} we present activity levels $V(t|G_i)$ for $G_0$ through $G_3$ for
all 16 events.



\end{document}